\UseRawInputEncoding
\documentclass[11pt]{article}
\usepackage[ margin=1in]{geometry}
\usepackage[affil-it]{authblk}
\usepackage{setspace}
\usepackage{tocloft}

\usepackage[sort&compress,numbers]{natbib}

\usepackage[font=small]{caption}
\usepackage{graphicx}
\usepackage{amsmath,amssymb,bm, bbold}
\usepackage{xr}
\usepackage{color}
\usepackage[normalem]{ulem}
\usepackage{wasysym}
\usepackage{siunitx}
\usepackage{physics}
\usepackage{lineno}
\usepackage{gensymb}
\usepackage[T1]{fontenc}
\usepackage{booktabs}
\usepackage[section]{placeins}
\usepackage{textcomp}
\usepackage[colorlinks=true, linkcolor=black, urlcolor=blue, citecolor=blue]{hyperref}
\usepackage{verbatim}
\usepackage{blindtext}
\usepackage{comment}
\usepackage{placeins}
\usepackage{upgreek}
\usepackage{textgreek}

\usepackage{epstopdf}

\makeatletter

\newcommand{\Rmnum}[1]{\expandafter\@slowromancap\romannumeral #1@}

\newcommand{\moire}{moir\'e }
\newcommand{\Moire}{Moir\'e }

\newcommand{\Vxxw}{$V_{xx}^\omega$}
\newcommand{\Exxw}{$E_{xx}^\omega$}
\newcommand{\Vxxwsq}{($V_{xx}^\omega$)$^2$}
\newcommand{\Vxytw}{$V_{xy}^{2\omega}$}
\newcommand{\Exytw}{$E_{xy}^{2\omega}$}
\newcommand{\Vxxtw}{$V_{xx}^{2\omega}$}
\newcommand{\Vbynm}{V\,nm$^{-1}$}
\newcommand{\D}{$D/\epsilon_0$}

\newcommand{\Vxytwnorm}{$\frac{V_{xy}^{2\omega}}{(V_{xx}^{\omega})^2}$}
\newcommand{\Exytwnorm}{$\frac{E_{xy}^{2\omega}}{(E_{xx}^{\omega})^2}$}
\newcommand{\Vxytwnormeq}{\frac{V_{xy}^{2\omega}}{(V_{xx}^{\omega})^2}}

\newcommand{\Rxx}{$R_{xx}$}

\newcommand{\sxx}{$\sigma_{xx}$}
\newcommand{\sxxsq}{$\sigma_{xx}^2$}

\newcommand{\uvec}[1]{\boldsymbol{\hat{\text{#1}}}}

\makeatother
\def\be{\begin{equation}}
\def\ee{\end{equation}}
\def \bea{\begin{eqnarray}}
\def \eea{\end{eqnarray}}

\begin{document}
	\title{Quantum geometric moment encodes stacking order of \moire matter}
	
	\author[1$\dagger$]{Surat Layek}
	\author[1$\dagger$*]{Subhajit Sinha}
	\affil[1]{Department of Condensed Matter Physics and Materials Science, Tata Institute of Fundamental Research, Homi Bhabha Road, Mumbai 400005, India.}
	
	\author[2$\dagger$*]{Atasi Chakraborty}
	\affil[2]{Institut f\"{u}r Physik, Johannes Gutenberg Universit\"{a}t Mainz, D-55099 Mainz, Germany.}
	
	\author[1]{Ayshi Mukherjee}
	
	\author[1]{Heena Agarwal}
	
	\author[3]{Kenji Watanabe}
	\affil[3]{Research Center for Functional Materials, National Institute for Materials Science, 1-1 Namiki, Tsukuba 305-0044, Japan.}
	
	\author[4]{Takashi Taniguchi}
	\affil[4]{International Center for Materials Nanoarchitectonics, National Institute for Materials Science,  1-1 Namiki, Tsukuba 305-0044, Japan.}
	
	\author[5*]{Amit Agarwal}
	\affil[5]{Department of Physics, Indian Institute of Technology, Kanpur 208016, India.}
	
	\author[1*]{Mandar M. Deshmukh}
	
	\affil[$\dagger$]{\textnormal{These authors contributed equally to this work}}
	\affil[*]{\textnormal{subhajit.sinha@tifr.res.in, atasi.chakraborty@uni-mainz.de, amitag@iitk.ac.in, deshmukh@tifr.res.in}}
	
	\date{}
	\maketitle
	\newpage
	\setlength{\columnsep}{5cm}
\section*{Abstract}
		Exploring the topological characteristics of electronic bands is essential in condensed matter physics. \Moire materials featuring flat bands provide a versatile platform for engineering band topology and correlation effects. In \moire materials that break either time-reversal symmetry or inversion symmetry or both, electronic bands exhibit Berry curvature hotspots. Different stacking orders in these materials result in varied Berry curvature distributions within the flat bands, even when the band dispersion remains similar. However, experimental studies probing the impact of stacking order on the quantum geometric quantities are lacking. 1.4$^\circ$ twisted double bilayer graphene (TDBG) facilitates two distinct stacking orders (AB-AB, AB-BA) and forms an inversion broken \moire superlattice with electrically tunable flat bands. The valley Chern numbers of the flat bands depend on the stacking order, and the nonlinear Hall (NLH) effect distinguishes the differences in Berry curvature dipole (BCD), the first moment of Berry curvature. The BCD exhibits antisymmetric behavior, flipping its sign with the polarity of the perpendicular electric field in AB-AB TDBG, while it displays a symmetric behavior, maintaining the same sign regardless of the electric field's polarity in AB-BA TDBG. This approach electronically detects stacking-induced quantum geometry, while opening a pathway to quantum geometry engineering and detection.
	
	\section{Introduction}
	Twistronics has emerged as a burgeoning field to engineer symmetry-broken flat bands that can be tuned electrically and via other knobs~\cite{adak_tunable_2024}.
	For example, magic-angle twisted bilayer graphene hosts a plethora of tunable correlated phases such as superconductivity~\cite{cao_unconventional_2018,lu_superconductors_2019} and orbital ferromagnetism~\cite{sharpe_emergent_2019-1,serlin_intrinsic_2020}.
	Recent advances in the field have drawn specific connections between electronic correlations in flat-band systems and the underlying band topology.
	For instance, the superconductivity and superfluidity in the flat bands of twisted multilayer graphene systems are known to arise from the quantum geometry of the flat bands~\cite{torma_superconductivity_2022}.
	It is also believed that fragile phases such as the fractional quantum anomalous Hall states~\cite{cai_signatures_2023, lu_fractional_2024} are better stabilized in bands with uniform Berry curvature~\cite{xie_fractional_2021} and high Chern numbers~\cite{herzog-arbeitman_moire_2024}. 
	As a result, the topology of the flat bands can provide important information not only on the Berry curvature distribution but also on the accompanying correlated phases it is susceptible to host.
	
	In this regard, twisted multilayer systems provide us with an additional knob to stack the multilayers with different stacking orders having distinct band topology.
	In some heterostructures, the stacking order leaves an imprint on the Berry curvature structure of the flat bands while keeping the energy dispersion of the bands similar.
	Engineering and studying such systems can help us determine the effects of the distinct topology of the bands on electronic transport.
	In addition, a change in stacking order across domain boundaries can induce unique topological electronic modes~\cite{ju_topological_2015}.
	Recently, domain boundaries across AB and BA domains in marginally twisted bilayer graphene have also been shown to host superconducting channels in the quantum Hall regime, highlighting the importance of studying the topology of distinct stacking orders of a system~\cite{barrier_one-dimensional_2024}.
	
	In this work, we explore the stacking order-induced differences in band topology by measuring the nonlinear Hall transport in twisted double bilayer graphene (TDBG).
	Owing to the \moire periodicity in TDBG, the $K$ and the $K'$ \moire bands decouple.
	This decoupling allows a valley Chern number C$_K$ (C$_{K'}$) to be defined for each \moire band of $K$ ($K'$) valley~\cite{song_topological_2015}.
	A nonzero C$_K$ (C$_{K'}$) quantifies the nontrivial topology of the $K$ ($K'$) \moire bands.
	In particular, the topological flat bands in TDBG~\cite{burg_correlated_2019,shen_correlated_2020,sinha_bulk_2020-4,cao_tunable_2020,liu_tunable_2020-1,wang_bulk_2022} have non-zero valley Chern numbers that depend on the stacking order--AB-AB or AB-BA.
	Tuning the valley Chern number, for example, via a perpendicular electric field~\cite{zhang_nearly_2019, koshino_band_2019-1,adak_perpendicular_2022}, corresponds to changing the Z$_2$~[$=(C_K-C_{K'})/2$] topology of the system.
	Recent experiments~\cite{sinha_berry_2022} and theoretical calculations~\cite{facio_strongly_2018,hu_nonlinear_2022, chakraborty_nonlinear_2022} have demonstrated that the Berry curvature dipole (BCD) senses topological transitions of the valley Chern type.
	Specifically, the BCD sign changes rapidly across specific topological Z$_2$ transitions~\cite{sinha_berry_2022, facio_strongly_2018}.
	Here, using nonlinear Hall measurements at zero magnetic field, we study the effect of stacking order on the BCD of flat bands.
	We demonstrate that experimentally probing the BCD variation across valley Chern transitions can distinguish the stacking order induced distinct band topology in differently stacked heterostructures.
	We vary the polarity of the perpendicular electric field and find that the Berry curvature, and hence the BCD evolves differently depending on the stacking order of $\approx1.4^\circ$ TDBG.
	Our experiments show that nonlinear Hall transport can be utilized to detect the distinct stacking-order induced BCD.
	
	\section{Results and Discussion}
	\subsection{Band Structure Calculations of Twisted Double Bilayer Graphene}
	\begin{figure*}[!h]
		\centering
		\includegraphics[width=16cm]{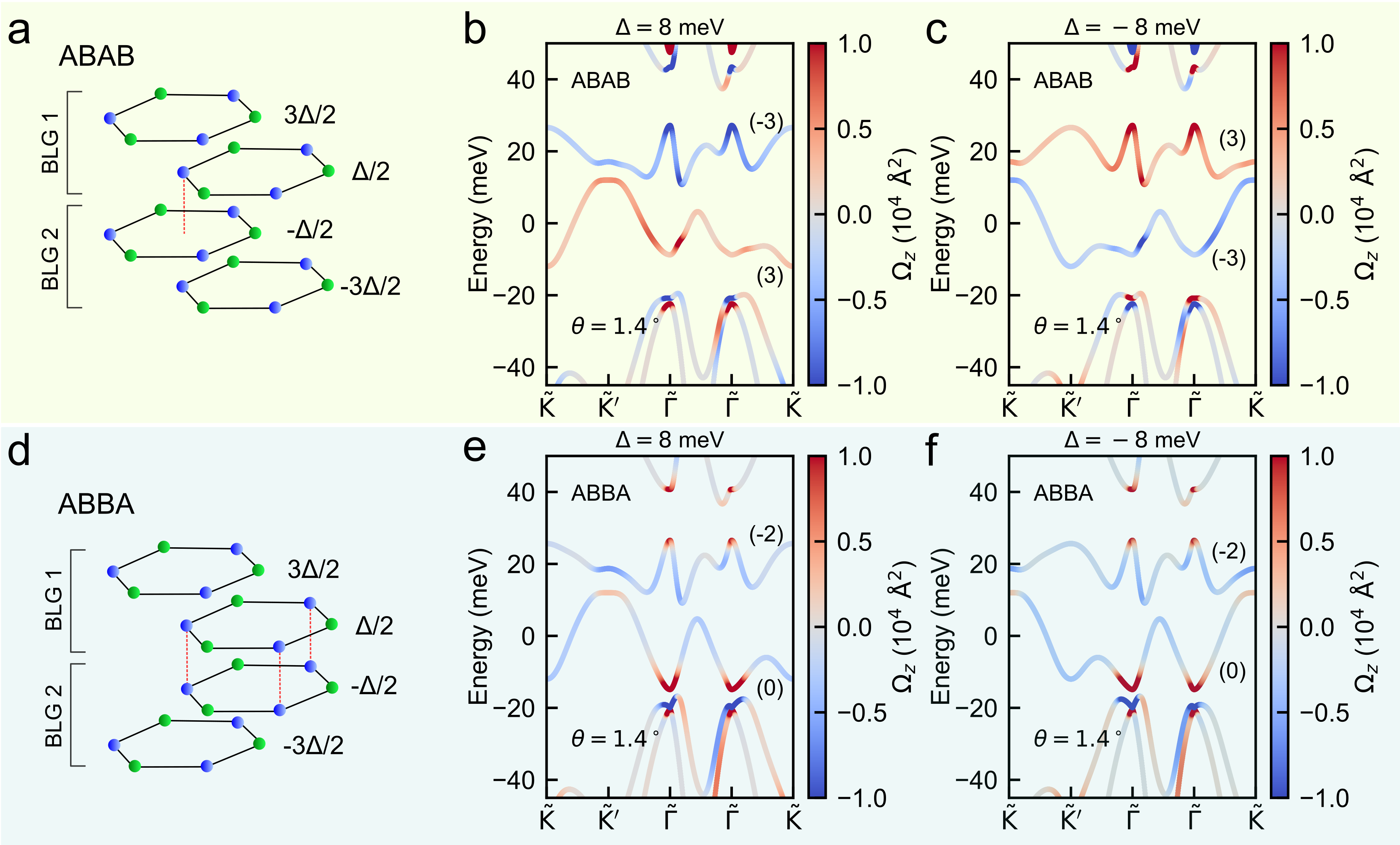}
		\caption{ \label{fig:fig1_theory} { \textbf{Change in Berry curvature distribution with perpendicular electric field in two distinct stacking orders of twisted double bilayer graphene (TDBG).}
				\textbf{a, d,}~Atomic arrangement in AB-AB (\textbf{a}) and AB-BA (\textbf{d}) stacked double bilayer graphene, before introducing any twist between the layers.
				The blue and green colors indicate the two different sublattices A and B.
				The dashed red line in \textbf{a} indicates that a sublattice of the top layer in BLG-2 lies at the hexagon center of the bottom layer in BLG-1 in the AB-AB arrangement.
				In \textbf{d}, the sublattices of the top layer in BLG-2 and the bottom layer in BLG-1 are aligned on top of each other.
				\textbf{b, c,}~Band structure of 1.40$^\circ$ AB-AB stacked TDBG for $\Delta=8$~meV (\textbf{b}) and $\Delta=-8$~meV (\textbf{c}).
				The twist angle $\theta$ in TDBG is introduced between the two bilayers, BLG-1 and BLG-2.
				\textbf{e, f,}~Band structure of 1.40$^\circ$ AB-BA stacked TDBG for $\Delta=8$~meV (\textbf{e}) and $\Delta=-8$~meV (\textbf{f}).
				The color indicates the Berry curvature ($\Omega_z$) of the bands.
				The valley Chern numbers are labeled for the flat bands.
				The $\Omega_z$ of the flat bands flip sign across \textbf{b} and \textbf{c}, while it remains of the same sign across \textbf{e} and \textbf{f}.
		}}
	\end{figure*}
	In TDBG, a Bernal (AB) bilayer graphene is stacked on another with a relative twist angle between them.
	Depending on how the second bilayer graphene is stacked (at an interlayer angle of $\theta$ or $180^{\circ}+\theta$), TDBGs have two predominant stacking orders: AB-AB-stacked~(Fig.~\ref{fig:fig1_theory}a) TDBG~\cite{adak_tunable_2020_prb}, or AB-BA-stacked~(Fig.~\ref{fig:fig1_theory}d) TDBG (see Supplementary Information Section~III for details on device fabrication).
	Our approach to distinguish how distinct stacking orders influence the band topology of TDBG is to pre-determine the band structure for a particular twist angle and identify characteristic differences in Berry curvature and BCD.
	We calculate the band structure of 1.4$^\circ$ AB-AB (Fig.~\ref{fig:fig1_theory}b, c) and AB-BA (Fig.~\ref{fig:fig1_theory}e, f) TDBG for positive ($\Delta=8$~meV) and negative ($\Delta=-8$~meV) interlayer potentials ($\Delta$).
	We note three observations. i) For a fixed $\Delta$, the valley Chern numbers of the flat bands are different for the two stacking orders, although the band dispersion is similar~\cite{koshino_band_2019-1}.
	ii) As we flip the polarity of $\Delta$, the sign of the Berry curvature distribution in the flat bands of the AB-AB TDBG flips (Fig.~\ref{fig:fig1_theory}b, c), whereas it remains unchanged in AB-BA TDBG (Fig.~\ref{fig:fig1_theory}e, f).
	At the phenomenological level, the Berry curvature sign flip in AB-AB TDBG is similar to the band inversion in AB bilayer graphene with $\Delta$ varying across $\Delta=0$ (we discuss this aspect later in Fig.~\ref{fig:fig5_BLG}). 
	iii) In the presence of time-reversal symmetry, the valley Chern numbers $C_K$ and $C_{K'}$ are equal and opposite, resulting in a total Chern number ($C=C_K+C_{K'}=0$) of zero, precluding any Berry curvature-driven linear anomalous Hall response.
	This prompts a natural question: Can we distinguish the band topology of these two stacking orders in transport experiments.
	To address this, in the following we present linear and nonlinear transport experiments (backed by theoretical calculations) that probe the BCD in AB-AB- and AB-BA-stacked TDBG, as perpendicular electric field switches polarity.
	
	\subsection{Linear and Nonlinear Hall transport}
\begin{figure*}[h]
		\centering
		\includegraphics[width=16cm]{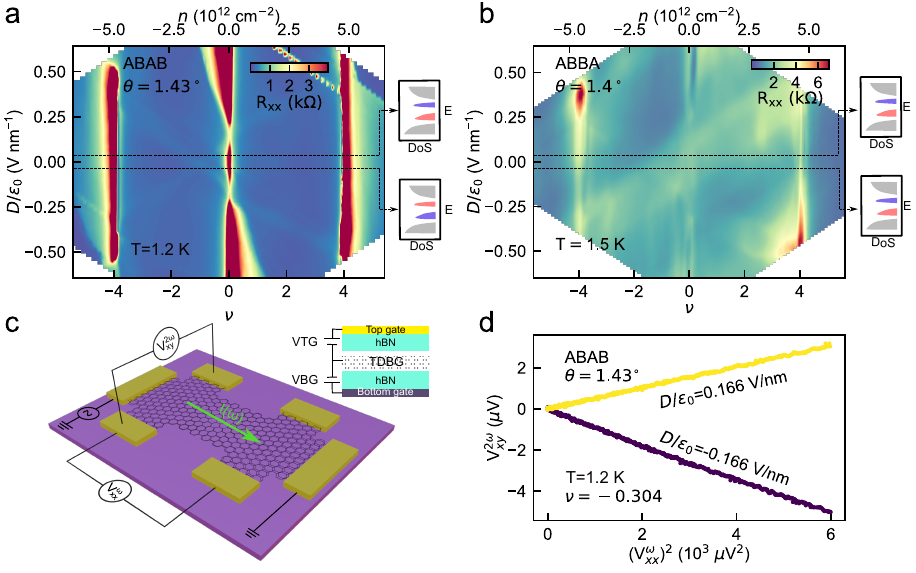}
		\caption{ \label{fig:fig2_Rxx} \textbf{Transport characterization of the two stacking orders, AB-AB and AB-BA, in TDBG.}
			\textbf{a, b,}~Longitudinal resistance \Rxx{} of 1.43$^\circ$ AB-AB~(\textbf{a}) and 1.40$^\circ$ AB-BA~(\textbf{b}) stacked TDBG devices as a function of filling factor ($\nu$) and perpendicular electric field (\D).
			The top axis in \textbf{a, b} indicates the charge density ($n$).
			The insets to the right in \textbf{a}, \textbf{b} are schematic representations to indicate the corresponding energy (E) vs density of states (DoS), close to the $K$ valley, for two polarities of \D. The colors indicate a nonzero Berry curvature of the flat bands that flip (does not flip) sign as the polarity of $D$ is reversed in $\approx1.4^\circ$ AB-AB (AB-BA) TDBG.
			The measurement temperature $T$ for \textbf{a} and \textbf{b} were 1.2 K and 1.5 K, respectively.
			\textbf{c,}~Measurement schematic for nonlinear Hall (NLH) voltage. An AC current, I($\omega$), is applied along the longitudinal direction of the device. The nonlinear Hall voltage, \Vxytw{} at twice the driving frequency (2$\omega$) and the longitudinal voltage, \Vxxw, at the driving frequency ($\omega$) are measured simultaneously while tuning the gate voltages to control the carrier density ($n$) and perpendicular electric field (\D{}). The inset shows the cross-sectional structure of the dual-gated device, consisting of the TDBG layer encapsulated between hexagonal boron nitride (hBN) layers, with independent top and bottom gate electrodes.
			\textbf{d,}~Variation of \Vxytw{} with \Vxxwsq{} for a fixed filling factor $\nu=-0.304$, for two different polarities of the perpendicular electric field \D{} in AB-AB TDBG. 
			A linear behavior of $V_{xy}^{2\omega}$ with ($V_{xx}^{\omega}$)$^2$ verifies the quadratic dependence of $V_{xy}^{2\omega}$ on current $I$($\omega$).
		}
	\end{figure*}

\begin{figure*}[!h]
		\centering
		\includegraphics[width=16cm]{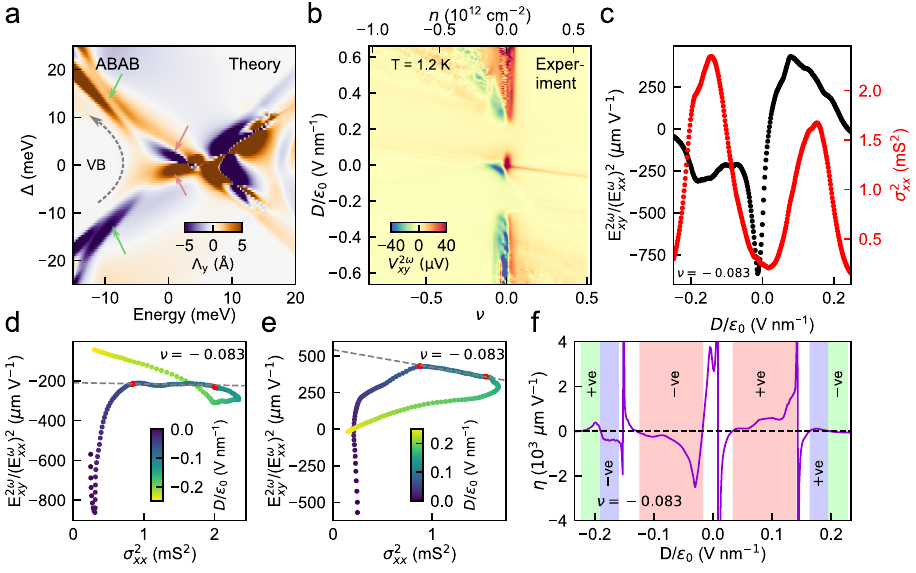}
		\caption{ \label{fig:fig3_ABAB} \textbf{Sign flip in BCD with change in \D{} polarity for 1.43$^\circ$ AB-AB TDBG.}
			\textbf{a,} Calculated y-component of BCD ($\Lambda_y$) as a function of energy and interlayer potential ($\Delta$) for 1.4$^\circ$ AB-AB TDBG.
			The calculation was performed with an uniaxial strain of 0.2\%, applied along the zig-zag axis of a BLG in TDBG.
			The dashed arrow is a guide to the eye that traces the movement of the valence band in energy with $\Delta$.
			The solid arrows show BCD sign changes when the polarity of $\Delta$ is reversed.
			\textbf{b,} Nonlinear Hall voltage (\Vxytw) as a function of filling factor ($\nu$) (corresponding $n$ is shown on the top-axis) and perpendicular electric field (\D) for 1.43$^\circ$ AB-AB twisted TDBG.
			\textbf{c,} \Exytwnorm{} (left axis; black data points) and \sxxsq{} (right axis; red data points) as a function of \D{} for a filling $\nu=-0.083$ in the valence band.
			\textbf{d, e} \Exytwnorm{} as a function of \sxxsq{} at $\nu=-0.083$, where \D{} is varied parametrically for $D<0$ (d) and $D>0$ (e).
			The red dots indicate the \D{} range within which the linear fit of the form \Exytwnorm=$\zeta$\sxxsq +$\eta$ is performed ($-$0.066 \Vbynm{} to $-$0.116 \Vbynm{} in \textbf{d}, and 0.078 \Vbynm{} to 0.124 \Vbynm{} in \textbf{e}). 
			The intercept $\eta$ changes sign across $\textbf{d}$ and $\textbf{e}$ as $D$ changes sign.
			\textbf{f,} The extracted local intercept $\eta$ as a function of \D{} for $\nu=-0.083$.
			The colors indicate the different \D{} ranges where $\eta$ flips sign across $D=0$.
			This captures the sign change of BCD on reversing the polarity of $D$ in the AB-AB TDBG.
			The measurements were performed at $T=1.2$~K.
		}
	\end{figure*}

	In Fig.~\ref{fig:fig2_Rxx}a and Fig.~\ref{fig:fig2_Rxx}b, we show the measured longitudinal resistance $R_{xx}$ as a function of the filling factor $\nu=4n/n_\text{S}$ and perpendicular electric field \D{} (the dual-gated geometry in our devices allow independent control of the charge density $n$ and \D, see Supplementary Section~IV.1 for details) for AB-AB TDBG and AB-BA TDBG, respectively.
	Here, $n_S=4.80\times10^{12}\ \mathrm{cm^{-2}}$ ($n_S=4.62\times10^{12}\ \mathrm{cm^{-2}}$) is the charge density required to fill or empty a flat band completely in AB-AB (AB-BA) TDBG.
	The twist angles of the two distinct stacking orders (see Supplementary Section~IV.1 for the twist angle estimations), are 1.43$^\circ$ (AB-AB) and 1.40$^\circ$ (AB-BA)
	(see Supplementary Section~VIII.2 for AB-AB TDBG device-2 with a twist angle of 1.1$^\circ$).
	The high values of $R_{xx}$ at $n=\pm n_S$ indicate the presence of \moire gaps.
	For fillings close to $\nu=0$, the $R_{xx}$ shows a minimum as $|D|/\epsilon_0$ is increased in both AB-AB and AB-BA TDBG, corresponding to a peak in conductivity squared \sxxsq{} in Fig.~\ref{fig:fig3_ABAB}c and Fig.~\ref{fig:fig4_ABBA}c, respectively.
	Such a feature is attributed to a gap closing and reopening transition~\cite{liu_tunable_2020-1} at a nonzero \D{} (see Supplementary Section~V for the temperature dependence of $R_{xx}$ at the charge neutrality gap) and is also reproduced in our theoretical calculations (see Supplementary Fig.~S3 for a band touching and reopening transition).
	The subtle differences in strain and twist angle of the two devices can possibly cause a difference in the measured value of the conductivity across the two devices.
	
	Recently, there has been a growing interest in studying the nonlinear effects in materials, owing to their connection with the quantum geometry of bands~\cite{ma_topology_2021,du_nonlinear_2021,BCD_Weyl_advmat}.
	In the presence of time-reversal symmetry, broken inversion symmetry is essential for nonzero Berry curvature.
	In multilayer systems such as bilayer graphene, the perpendicular electric field breaks the inversion symmetry and introduces a nonzero Berry curvature at the band edge.
	A broken C$_3$ symmetry (such as due to non-zero in-plane strain in \moire superlattices~\cite{kazmierczak_strain_2021,mcgilly_visualization_2020, li_unraveling_2021}, see Supplementary Section~IV.2 for evidence of strain in our TDBG device) together with broken inversion symmetry, creates a non-uniform Berry curvature distribution in $k$-space resulting in a nonzero BCD, $\Lambda_{\alpha} = \sum_{n} \int_{\rm mBZ}  \dfrac{d{\bf k}}{(2\pi)^2} \Omega_z^n \frac{\partial\epsilon^n_{\bf k}}{\hbar \partial k_{\alpha}} \frac{\partial f(\epsilon^n_{\bf k})}{\partial \epsilon^n_{\bf k}}$.
	Here, the integral is carried over the \moire Brillouin zone (mBZ), $\alpha$ stands for the spatial index $(x,y)$, $\epsilon_{\bf k}^n$ is the energy of the $n^{th}$ band, $f(\epsilon^n_{\bf k})$ is the Fermi-Dirac function, and a sum over all the bands crossing the Fermi energy is implied.
	A nonzero BCD generates a second-order nonlinear Hall response $\vec{j}^{2\omega} \propto \uvec{z} \cross \vec{E}^\omega(\vec{\Lambda} \cdot \vec{E}^\omega)$ that is detected by measuring~\cite{fu_quantum_2015} the nonlinear Hall (NLH) voltage \Vxytw{}.
	Figure~\ref{fig:fig2_Rxx}c shows our schematic to measure the NLH voltage. 
	The linear dependence of \Vxytw{} on \Vxxwsq{} in Fig.~\ref{fig:fig2_Rxx}d confirms the characteristic second-order nature of the measured \Vxytw{} in the AB-AB TDBG device (see Supplementary Section~VI for additional characterization of nonlinear voltage in the TDBG devices).
	NLH response has been investigated in transition metal dichalcogenides (TMDCs)~\cite{kang_nonlinear_2019,ma_observation_2019,shvetsov_nonlinear_2019,xiao_berry_2020,tiwari_giant_2021,kumar_room-temperature_2021,son_strain_2019,huang_giant_2022_published}, corrugated graphene~\cite{ho_hall_2021}, 3D systems~\cite{facio_strongly_2018,sun_berry_2018}, and recently in few \moire superlattices owing to both BCD~\cite{sinha_berry_2022, huang2023intrinsic, datta_nonlinear_2024} and scattering~\cite{duan_giant_2022, he_graphene_2022} mechanisms.
	Hence it is important to devise a pathway forward to systematically analyze and segregate the intrinsic and extrinsic mechanisms.
	Next, we systematically compare the measured \Vxytw{} vs. \D{} dependence across a change in the polarity of \D{}, which distinguishes the band topology of AB-AB and AB-BA TDBG.
	
	\subsection{Berry Curvature Dipole calculations and Scaling Analysis}
	In Fig.~\ref{fig:fig3_ABAB}a, we show the calculated $\Lambda_y$ for 1.4$^\circ$ AB-AB TDBG as a function of energy and $\Delta$ (where $\Delta$ is proportional to \D; see Supplementary Fig.~S4 for the BCD dependence on a greater energy range).
	The choice of a twist angle of 1.4$^\circ$ allows us to explore the BCD of isolated flat bands.
	We see that as $\Delta$ is flipped, $\Lambda_y$ changes its sign.
	The sign reversal is most apparent for the valence band.
	In Supplementary Section~I (Fig.~S1 and Fig.~S2), we show the band structure calculations with a nonzero strain and plot the corresponding BCD vs energy lineslices for different $\Delta$.
	As $\Delta$ is varied and flipped, the flat bands undergo band touchings and consequently, the Berry curvature distribution and valley Chern numbers change reflecting in the sign change of BCD.
	
	To experimentally detect this sign reversal of BCD, we measure \Vxxw{} (Fig.~\ref{fig:fig2_Rxx}a shows the corresponding \Rxx=\Vxxw/$I$, where $I$ is the channel current) and \Vxytw{} (Fig.~\ref{fig:fig3_ABAB}b) as a function of the perpendicular electric field \D{} and fillings close to the charge neutrality point $\nu=0$.
	\Vxxw{} and \Vxytw{} correspond to the linear \Exxw{}(=\Vxxw/$L$) and nonlinear \Exytw{}(=\Vxytw/$w$) in-plane electric fields, where $L$ and $w$ are the length and width of the device, respectively. 
	In general, the measured \Vxytw{} contains the intrinsic BCD contribution along with extrinsic contributions such as the skew scattering and side-jump mechanisms.
	A way forward to segregate the intrinsic BCD contribution from other extrinsic contributions is to study the linear scaling of the form~\cite{du_disorder-induced_2019, sinha_berry_2022,kang_nonlinear_2019} \Exytwnorm=$\zeta$\sxxsq +$\eta$ (over a small window of \D{}), where $\zeta$ and $\eta$ are the slope and intercept, respectively (see Supplementary Section~VII.1 for details).
	Here, the intercept $\eta$ is used as an order of magnitude estimation~\cite{kang_nonlinear_2019} of BCD$\sim \eta E_F/e$, where $E_F$ is the Fermi energy.
	Figure~\ref{fig:fig3_ABAB}c shows a representative lineslice of the \Exytwnorm{} and \sxxsq{} with \D, for a fixed filling of $\nu=-0.083$ in the valence band.
	Figure \ref{fig:fig3_ABAB}d and \ref{fig:fig3_ABAB}e probes the scaling relation for -ve and +ve values of $D$, respectively. 
	Here, $D$ is varied as a parameter to probe the linear scaling relation (see Supplementary Section~VII.2).
	We first probe the scaling for both polarities of the perpendicular displacement field $D$ for $|D|<|D^*|$, where \sxx{} is maximum at $|D^*|/\epsilon_0 \approx 0.16$~\Vbynm{} corresponding to the gap closing discussed earlier.
	We find that the intercept $\eta$ changes sign when fitted linearly within a similar $|D|/\epsilon_0$ range across Fig.~\ref{fig:fig3_ABAB}d and Fig.~\ref{fig:fig3_ABAB}e.
	This choice of $|D|/\epsilon_0$ range guarantees that the analysis is performed in a $|D|/\epsilon_0$ range within which no drastic band structure changes such as a gap closing and reopening occurs.
	Although our devices at this twist angle show non-zero \Vxxtw{} (see Supplementary Section~VI) that is typically attributed to extrinsic scattering mechanisms~\cite{he_graphene_2022}, a sign change in intercept ($\eta$) with \D{} cannot be explained via scattering mechanisms alone~\cite{ma_observation_2019}.
	The sign change in intercept ($\eta$) agrees with our calculated BCD sign reversal with the polarity of $\Delta$ in the valence band (Fig.~\ref{fig:fig3_ABAB}a), and captures the intrinsic contribution at this filling (see Supplementary Section~VIII.1 for similar results at other fillings).
	
	\begin{figure*}[!h]
		\centering
		\includegraphics[width=16cm]{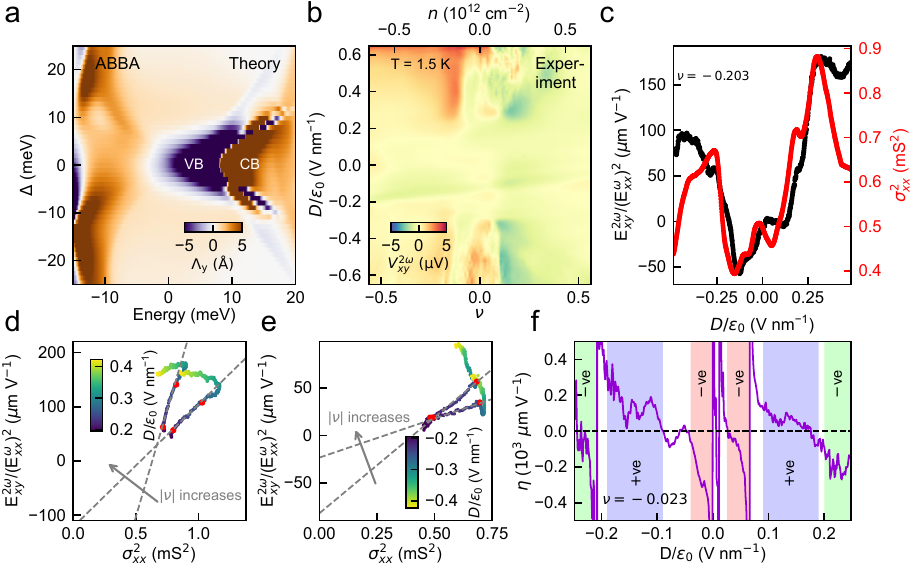}
		\caption{ \label{fig:fig4_ABBA} \textbf{Sign of BCD is intact with change in \D{} polarity for 1.40$^\circ$ AB-BA TDBG.}
			\textbf{a,} Calculated y-component of BCD ($\Lambda_y$) as a function of energy and interlayer potential ($\Delta$) for 1.40$^\circ$ AB-BA TDBG.
			The calculation was performed with the same strain parameters as in Fig.~\ref{fig:fig3_ABAB}a.
			\textbf{b,} Nonlinear Hall voltage (\Vxytw) as a function of filling factor ($\nu$) (corresponding $n$ is shown on the top-axis) and perpendicular electric field (\D) for 1.40$^\circ$ AB-BA twisted TDBG at $T= 1.5$~K.
			\textbf{c,} \Exytwnorm{} (left axis; black data points) and \sxxsq{} (right axis; red data points) as a function of \D{} for a filling $\nu=-0.203$ in the valence band.
			\textbf{d, e} \Exytwnorm{} as a function of \sxxsq{} where \D{} is varied parametrically for $D>0$ (d) and $D<0$ (e) for the fixed filling factors $\nu=-0.203,-0.248$ in the valence band.
			The dashed gray line indicates a linear fit of the form \Exytwnorm=$\zeta$\sxxsq +$\eta$ for the two fillings and the red dots indicate the fitting range.
			The intercept $\eta$ does not change sign across $\textbf{e}$ and $\textbf{f}$ even though $D$ changes sign.
			\textbf{f,}~The extracted local intercept $\eta$ as a function of \D{} for $\nu=-0.023$.
			The colors indicate the different \D{} ranges where $\eta$ does not flip sign across $D=0$.
			This captures the fact that the BCD does not change sign in the AB-BA TDBG on reversing the polarity of $D$.
			The measurements were performed at $T=1.5$~K.
		}
	\end{figure*}
	
	We now focus on how reversing the polarity of \D{} affects the BCD sign, for an extended \D{} range.
	We probe the local intercept $\eta$ as a function of \D{} (here, $\eta$ is defined locally for a small moving window of \D{}; see Supplementary Section~VII.3 for details of the analysis) across band-touching transitions in Fig.~\ref{fig:fig3_ABAB}f.
	The intercept $\eta$ changes sign for the similar magnitude range of \D{} but opposite polarity, indicated by the same color.
	The opposite signs of intercepts for the opposite polarity of \D{} indicate that the BCD flips sign once the electric field polarity is reversed in AB-AB TDBG (Supplementary Fig.~S4b,c shows that the dependence of the valley Chern number with $\Delta$ is anti-symmetric).
	Interestingly, we also see a sharp change in the intercept $\eta$ in the white-colored regions across valley Chern transitions.
	The BCD is theoretically known to increase and switch rapidly across a band touching topological transition~\cite{facio_strongly_2018}, which further confirms the intrinsic-dominated origin of the measured $\eta$.
	We next carry out the same analysis for a 1.4$^\circ$ AB-BA TDBG device to examine the BCD evolution when the polarity of the perpendicular electric field flips.
	
	In Fig.~\ref{fig:fig4_ABBA}a, we show the theoretically calculated BCD, $\Lambda_y$, in the flat bands of 1.4$^\circ$ AB-BA TDBG.
	In contrast to 1.4$^\circ$ AB-AB TDBG discussed in Fig.~\ref{fig:fig3_ABAB}a, we do not see a sign change in the calculated BCD as the polarity of $\Delta$ is reversed. 
	This is analogous to an AA-bilayer graphene system (discussed later in Fig.~\ref{fig:fig5_BLG}c, d).
	To experimentally validate this observation, we measured the \Vxytw{} in AB-BA TDBG with a twist angle of 1.40$^\circ$ at 1.5 K in Fig.~\ref{fig:fig4_ABBA}b.
	Figure~\ref{fig:fig4_ABBA}c shows a representative lineslice of the \Exytwnorm{} and \sxxsq{} with \D, for a fixed filling of $\nu=-0.203$ in the valence band to test the scaling relation.
	Figure \ref{fig:fig4_ABBA}d and \ref{fig:fig4_ABBA}e probe the scaling relation for +ve and -ve values of $D$, respectively, where $D$ is varied as a parameter for two fixed fillings in the valence band (see Supplementary Section~IX for other fillings $\nu$).
	Interestingly, in this case, the intercept $\eta$ does not change sign across a change in $D$ polarity (Fig.~\ref{fig:fig4_ABBA}d and \ref{fig:fig4_ABBA}e), in agreement with the BCD calculation presented in Fig.~\ref{fig:fig4_ABBA}a.
	The decrease in intercept $\eta$ with increasing |$\nu$| placed inside the flat valance band further agrees with AB-BA TDBG studied in Zhong et al.~\cite{zhong_effective_2024}, and indicates the domination of intrinsic contribution in this $\nu$ range.
	The fact that the BCD in the AB-BA TDBG does not flip with a change in polarity of $D$ is also evident in the dependence of the local intercept $\eta$ over an extended \D{} range in Fig.~\ref{fig:fig4_ABBA}f (Supplementary Fig.~S4e,f shows that the dependence of the valley Chern number with $\Delta$ is symmetric for 1.4$^\circ$ AB-BA TDBG).

	\subsection{Nonlinear Hall transport in Bilayer graphene}
\begin{figure*}[!h]
		\centering
		\includegraphics[width=16cm]{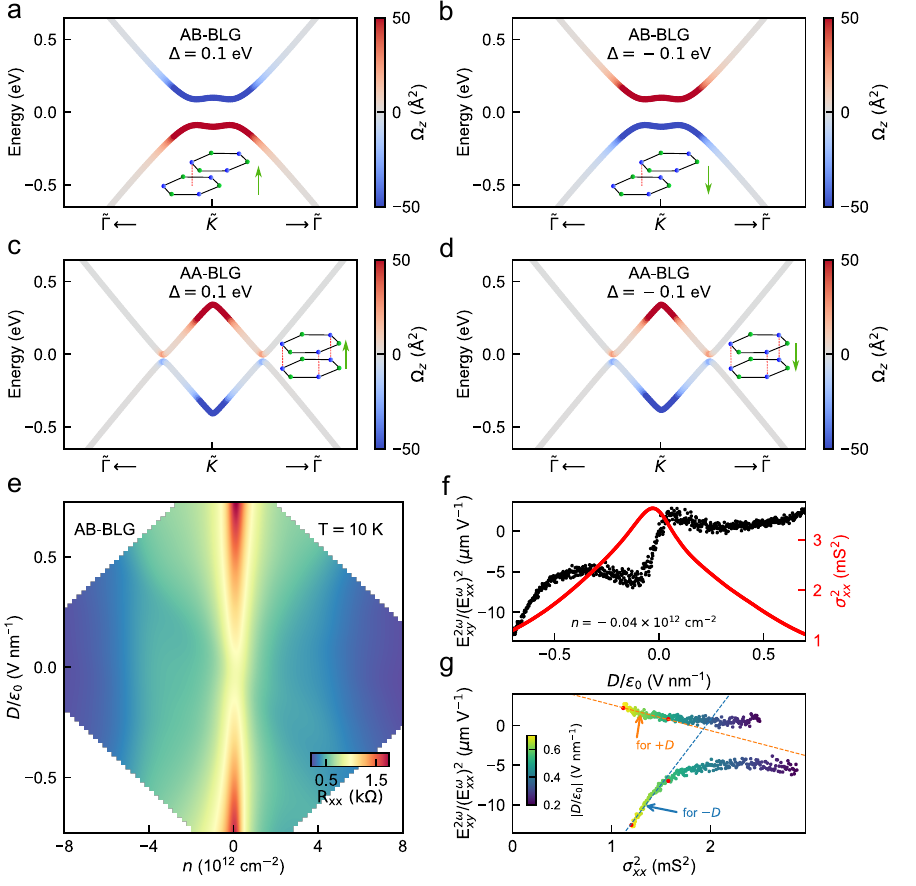}
		\caption{ \label{fig:fig5_BLG} \textbf{BCD sign reversal in AB-stacked bilayer graphene.}
			\textbf{a, b,}~Band structure of AB-bilayer graphene for $\Delta=0.1$~eV (a) and $\Delta=-0.1$~eV (b). 
			Insets show the atomic arrangement of Bernal (AB) bilayer graphene, where the green arrow indicates the direction of applied interlayer potential $\Delta$ (a nonzero $\Delta$ translates to an applied perpendicular \D{} in experiments).
			The Berry curvature of each band (indicated by color) changes sign as the sign of $\Delta$ (polarity of $D$) is flipped.
			\textbf{c, d,}~Band structure of AA-bilayer graphene for $\Delta=0.1$~eV (c) and $\Delta=-0.1$~eV (d), with an induced gap (see Supplementary Section~II.1) to have nonzero Berry curvature. 
			Insets show the atomic arrangement of AA-stacked bilayer graphene, where the green arrow indicates the direction of applied $\Delta$.
			The Berry curvature of each band (indicated by color) does not change sign as the sign of $\Delta$ is flipped.
			\textbf{e,}~Longitudinal resistance \Rxx{} as a function of $n$ and \D{} in AB-stacked bilayer graphene.
			\textbf{f,}~\Exytwnorm{} (left axis; black data points) and \sxxsq{} (right axis; red data points) as a function of \D{} for a fixed charge density $n=-0.04 \cross 10^{12}$~cm$^{-2}$ close to the charge neutrality gap.
			\textbf{g,}~\Exytwnorm{} as a function of \sxxsq{} where \D{} is varied parametrically, for $D>0$ (upper plot) and $D<0$ (lower plot) for the fixed charge density $n=-0.04 \cross 10^{12}$~cm$^{-2}$.
			The dashed orange (blue) line indicates a linear fit of the form \Exytwnorm=$\zeta$\sxxsq +$\eta$, performed for the high positive (negative) \D{} range, and the red dots indicate the fitting range.
			The intercept $\eta$ is positive (negative) for $D>0$ ($D<0$), thus indicating that the BCD of AB-bilayer graphene changes sign as the perpendicular electric field is flipped.
			The measurements in \textbf{e-g} were performed at 10 K.
		}
	\end{figure*}

	A simpler system in which the BCD sign reversal with the reversal of $D$ polarity is expected, analogous to AB-AB TDBG discussed earlier in Fig.~\ref{fig:fig3_ABAB}, is the Bernal (AB-stacked) bilayer graphene (BLG).
	Figure~\ref{fig:fig5_BLG}a, b, shows the low-energy bandstructure of AB-BLG, where the Berry curvature, and consequently the BCD, of the bands flip with the polarity of $\Delta$ (see Supplementary Section~II.2 and Section~II.3 for details of theoretical calculations).
	To experimentally verify this BCD sign reversal, we fabricated a dual-gated bilayer graphene device.
	Figure~\ref{fig:fig5_BLG}e shows the measured \Rxx{} with $n$ and \D{}.
	We note that at the charge neutrality point ($n=0$), the \Rxx{} increases as the magnitude of the perpendicular electric field $|D|/\epsilon_0$ is increased.
	This is due to the band gap opening with \D{} at charge neutrality in BLG, consistent with earlier works~\cite{zhang_direct_2009}.
	In Fig.~\ref{fig:fig5_BLG}f, we plot the measured \Exytwnorm{} and \sxxsq{} as a function of \D{} close to the charge neutrality point.
	To extract the intrinsic BCD contribution at the band edge, in Fig.~\ref{fig:fig5_BLG}g we plot \Exytwnorm{} with \sxxsq{} parametrically as a function of \D{}, for both the polarities of \D.
	We fit the \Exytwnorm{} vs. \sxxsq{} dependence with the linear scaling relation used earlier, in the high \D{} regime where the variation of Berry curvature with \D{} is relatively low (see Supplementary Section II.4).
	We find that the intercept $\eta$ of the linear scaling (dashed line in Fig.~\ref{fig:fig5_BLG}g), and thus the BCD, indeed flips with a reversal in \D{} polarity, in analogy to the BCD sign reversal of AB-AB TDBG discussed in Fig.~\ref{fig:fig3_ABAB}.
	On the contrary, the calculated Berry curvature of AA-stacked BLG does not flip as the polarity of \D{} is flipped (Fig.~\ref{fig:fig5_BLG}c, d), analogous to ABBA-TDBG discussed in Fig.~\ref{fig:fig4_ABBA}.
	Together, the experimental observation along with the theoretical calculations on BLG demonstrate that the nonlinear Hall transport is sensitive to the BCD sign reversal with the \D{} polarity in AB-BLG.
	
	\section{Conclusion}
	In summary, we find that TDBG has two distinct stacking orders, namely AB-AB and AB-BA, with similar band dispersion but different valley Chern numbers, most apparent when the flat bands are isolated from the remote bands.
	The parameter space for tuning bands in TDBG is substantially large; it comprises of twist angle $\theta$, strain \%, and \D.
	In particular, \D{} tunes the valley Chern numbers of the flat bands in TDBG.
	We demonstrate a way to electrically distinguish the flat band quantum geometry of the two distinct stacking orders of $\approx$1.4$^\circ$ TDBG that have different valley Chern numbers, by studying the nonlinear Hall voltage as a function of the perpendicular electric field (\D).
	Our central observation is that the sign of BCD is odd as a function of the perpendicular electric field for one stacking (AB-AB) and even for the other stacking (AB-BA). 
	Our study offers an example of how the stacking of layers provides insight into the distinct topological structure of electronic bands, using the nonlinear Hall effect.
	Our work motivates the use of nonlinear Hall transport to probe and identify differently stacked twisted heterostructures, such as twisted transition metal dichalcogenides~\cite{ma_relativistic_2024}, or other 2D materials.
	
	\section*{Acknowledgements:}
	We thank Pratap Chandra Adak for his experimental assistance and helpful comments.
	We thank U. Chandni for the thoughtful discussions and comments.
	M.M.D. acknowledges the Department of Science and Technology (DST) of India for J.C. Bose fellowship 	JCB/2022/000045,  Nanomission grant SR/NM/NS45/2016, and DST SUPRA grant SPR/2019/001247, CEFIPRA CSRP Project no. 70T07-1 along with the Department of Atomic Energy of Government of India 12-R\&D-TFR-5.10-0100 for support.
	A.A thanks the Department of Science and Technology for Project No. DST/NM/TUE/QM-6/2019(G)-IIT Kanpur, of the Government of India, for financial support.
	K.W. and T.T. acknowledge support from the Elemental Strategy Initiative conducted by the MEXT, Japan (grant no. JPMXP0112101001), and JSPS KAKENHI (grant nos. 19H05790 and JP20H00354).
	A.C. acknowledges the A. V. Humboldt Foundation for financial support.
	We thank CC-IIT  Kanpur, for the high-performance computing facility.
	
	\section*{Conflict of Interest}
	The authors declare no conflict of interest.
	
	\section*{Author Contributions:}
	S.L. and S.S fabricated the devices. A.M. and H.A. helped in fabrication. S.S. and S.L. did the measurements and analyzed the data. A.C. and A.A. did the theoretical calculations. K.W. and T.T. grew the hBN crystals. S.S., S.L., A.C., and M.M.D. wrote the manuscript with inputs from all authors. M.M.D. supervised the project.
	
	\section*{Data Availability:}
	The data related to this study are available from the corresponding authors upon reasonable request.
	
	\section*{Code Availability:}
	The code that supports the findings of this study is available from the corresponding authors upon reasonable request.

\clearpage
\setcounter{section}{0}
\renewcommand{\thesection}{\Roman{section}}
\renewcommand{\thefigure}{S\arabic{figure}}
\captionsetup[figure]{labelfont={bf},name={Fig.}}
\setcounter{figure}{0}
\renewcommand{\theHequation}{Sequation.\theequation}
\renewcommand{\theequation}{S\arabic{equation}}
\setcounter{equation}{0}

\begin{center}
\Large{Supplementary Information} \\ \Large{Quantum geometric moment encodes stacking order of \moire matter}
\end{center}

\newpage
\doublespacing
\newpage

\section{Continuum model Hamiltonian of TDBG}
To construct the moir\'{e} Hamiltonian for different stacked TDBG platform, we first start with a brief review of the low energy model for Bernal stacked bilayer graphene building blocks. The Brillouin zone (BZ) of the AB-stacked bilayer graphene is identical to that of monolayer graphene. The primitive lattice vectors ${\bm a}_1 = a(1,0)$ and ${\bm a}_2 = a(1/2, \sqrt{3}/2)$ yield the reciprocal lattice vectors to be ${\bm b}_1^{(0)} = \frac{4 \pi}{\sqrt{3} a} (\sqrt{3}/2 , - 1/2)$ and ${\bm b}_2^{(0)} =\frac{4 \pi}{\sqrt{3} a} (0, 1)$. Here, $a$ is the lattice constant which is $\sqrt{3}$ times the carbon-carbon bond length $d=1.42$ \AA. The coordinates of the vertices of the hexagonal first BZ are ${\bm K}_{\xi}^{(0)} = \xi (2 {\bm b}_1^{(0)} + {\bm b}_2^{(0)})/3$ with $\xi=\pm 1$ being the valley index.

Including the effects of hexagonal warping, the Hamiltonian near the $K$-valley can be expressed in terms of the fermion operators of the $A$ and the $B$ sublattice of the top and the bottom layers, $[c^t_A(k), c^t_B(k), c^b_A(k), c^b_B(k)]$, as
\be \label{ham_blg}
H(k)=\begin{pmatrix}
h_k^t & t_k \\ t_k^\dagger & h_k^b
\end{pmatrix}.
\ee
Here, the block diagonal matrices $h_k^{t/b}$ represent the $2\times 2$ massive Dirac Hamiltonian of the top and bottom monolayers and $t_k$ represents the effect of inter-layer hopping. The corresponding matrices are
\be
h_k^{(t/b)} = 
\hbar v_0 {\bm \sigma} \cdot {\bm k}  + \frac{\delta}{2} ( \mathbb{I} \mp \sigma_z),~~ t_k= \begin{pmatrix}
-\hbar v_4 \pi^\dagger & -\hbar v_3 \pi \\ \gamma_1 & -\hbar v_4 \pi^\dagger
\end{pmatrix}~,
\ee
\noindent with $\pi\equiv k_x + i k_y$. In the Hamiltonian, different intra-layer and inter-layer couplings have been introduced through the hopping parameter $\gamma_i$ or equivalently by $v_i = \sqrt{3} |\gamma_i| a /(2 \hbar)$. The nearest neighbor intra-layer coupling between the $A$ and the $B$ sublattice is represented by the parameter $v_0$. The inter-layer intra-dimer coupling is represented through $\gamma_1$. Parameters $\gamma_3$ and $\gamma_4$ are the couplings between the inter-layer non-dimer sites and the inter-layer coupling between dimer and non-dimer sites, respectively. 
For our calculations we consider $\delta=15 $ meV, $\gamma_0 = -3.1$ eV, $\gamma_3=283$ meV and $\gamma_4=138$ meV. 

\begin{figure*}[h]
    \centering
    \includegraphics[width=16cm]{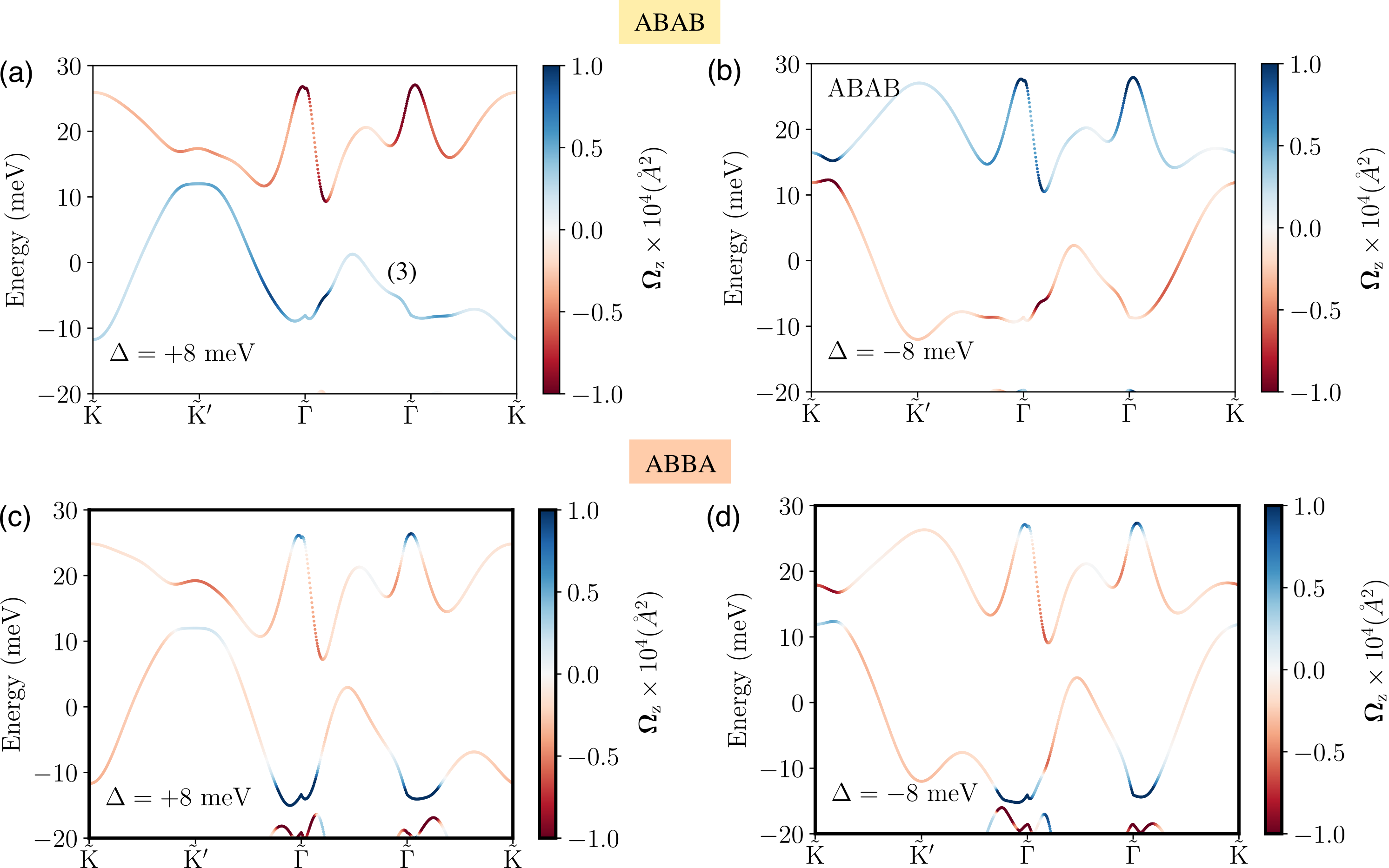}
    \caption{ \label{sfig:fig1_theory} \textbf{ Band dispersion along high-symmetry paths in presence of  $\epsilon=0.2\%$ strain.} (a) and (b) represent the Berry curvature color coded energy dispersion of AB-AB stacked TDBG for $\Delta=8$ meV and $\Delta=-8$ meV respectively. The Berry curvature for each band changes sign on reversing the direction of displacement field which is captured by the sign change of $\Delta$. Momentum-dependent Berry curvature resolved band-dispersion of AB-BA for $\Delta=\pm 8$ meV are shown in (c) and (d). The sign of Berry curvature is invariant with the applied electric field direction for AB-BA stacked TDBG. }
\end{figure*}

The reciprocal lattice vectors of the \moire lattice are obtained as ${\bm G}_{m}^i={\bm b}_i^{(1)}-{\bm b}_i^{(2)}$, with the rotated reciprocal lattice vectors of each bilayer being specified by ${\bm b}_1^{(l)} = {\cal R}(\mp \theta/2) {\bm b}_1^{(0)}$ with $\mp$ for bilayer $l=1,2$, respectively. Using this, we obtain the pair of primitive \moire lattice vectors to be ${\bm G}_{m}^1=\frac{8 \pi}{\sqrt{3}a} \sin \frac{\theta}{2}(-1/2, \sqrt{3}/2)$ and ${\bm G}_{m}^2=\frac{8 \pi}{\sqrt{3}a} \sin \frac{\theta}{2}(1/2, \sqrt{3}/2)$. Using the low energy Hamiltonian [Eq.~\eqref{ham_blg}] for each lattice point, vertices of the small moir\'{e} hexagons, and the moir\'e hopping matrix, we construct the continuum Hamiltonian for TDBG. A certain cut-off in the reciprocal space is used to truncate the lattice. 
The smallest TDBG AB-AB Hamiltonian for the $K$-valley can be written as~\cite{sinha_berry_2022_SI,Atasi2022T_SI,Chakraborty2022_SI,Dutta2023_SI}

\be 
{\mathcal H} = 
\begin{pmatrix}
	h_{k,t}^++\Delta_t^+ & t_k^{+} & 0 & 0 \\
	{t^+_k}^\dagger & h_{k,b}^{+}+\Delta_b^+ & T & 0 \\
	0 &  T^\dagger & h_{k,t}^-+\Delta_t^- &  t_k^-\\
	0 & 0 & {t_k^-}^\dagger  & h_{k,b}^{-}+\Delta_b^- \\
\end{pmatrix} .
\label{tdbg_ham}
\ee
Here, the superscripts on $h^{\pm}_{k,t/b}$ represents rotated Dirac Hamiltonian as $h^{\pm}={\mathcal R}(\mp \theta/2) {\bm k}\cdot{\bm \sigma}$ and $\Delta_{t/b}^{\pm}$ represents the effect of the perpendicular electric field. In Eq.~\eqref{tdbg_ham}, $T({\bm r})$ represents the moir\'e coupling matrix, which connects the bottom B-layer of bilayer-$1$ to the top layer of A-bilayer-$2$. For the AB-BA stacked he the double-bilayer graphene, the inter-layer hopping matrix T connects the bottom B-layer of bilayer-$1$ to the top layer of B-bilayer-$2$.

  In this smallest TDBG Hamiltonian, only the nearest neighbor coupling will be considered, which is connected by the vectors ${\bm q}_b= \frac{8\pi}{3a} \sin \frac{\theta}{2}(0, -1)$ and  ${\bm q}_{tl}  =\frac{8\pi}{3 a} \sin \frac{\theta}{2}(-\sqrt{3}/2, 1/2)$ and  ${\bm q}_{tr} =\frac{8\pi}{3 a} \sin \frac{\theta}{2}(\sqrt{3}/2, 1/2)$. The moir\'e hopping matrices are given by 
\be
T({\bm r}) =  \sum_{j=b, tr,tl}T_{{\bm q}_j} e^{-i{\bm q}_j \cdot {\bm r}}~,~~~{\rm where}
\ee
\be 
T_b = 
\begin{pmatrix}
\omega & \omega' \\
\omega' & \omega
\end{pmatrix}
~~~~~~~~T_{tr/tl} = 
\begin{pmatrix}
\omega & \omega' e^{\mp i 2\pi/3} \\
\omega' e^{\mp i 2\pi/3} & \omega
\end{pmatrix}~.
\ee
Here, $\omega$  and $\omega'$ denote the diagonal and the off-diagonal hopping strengths, respectively. We emphasize that an unequal $\omega$ and $\omega'$, specifically $\omega' > \omega$, is crucial to match the calculated low energy band structure with the experimentally observed spectral gap~\cite{koshino_band_2019-1_SI, chebrolu_PRB2019_flat_SI}. 
In this paper, we consider $\omega'=106$ meV and $\omega=79$ meV~\cite{koshino_band_2019-1_SI, mohan_PRB2021_tri_SI}.

\begin{figure*}[t]
    \centering
    \includegraphics[width=12cm]{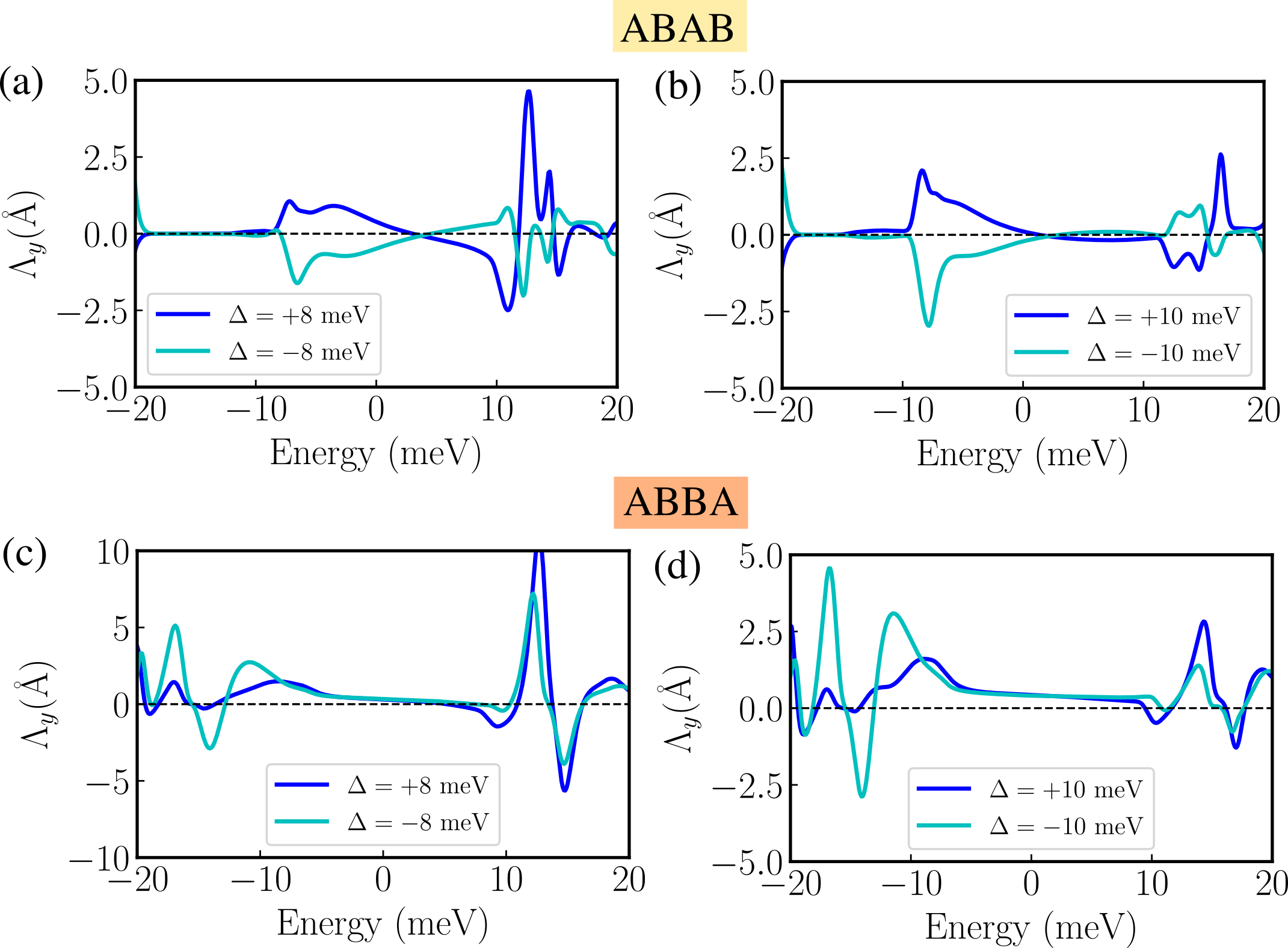}
    \caption{ \label{sfig:fig2_theory} \textbf{ Variation of Berry Curvature Dipole with the sign of electric field.} The line cut of BCD with the energy of AB-AB TDBG for $\pm$8 meV (a) and $\pm$10 meV (b). The BCD flips sign with opposite $\Delta$. (c) and (d) represent the BCD for $\pm$8 meV  and $\pm$10 meV similarly for AB-BA TDBG. Here the BCD peak-structure shows the same sign for opposite out-of-plane electric field values. }
\end{figure*}

\noindent In the presence of uni-axial strain ($\mathcal{E}$), the Dirac Hamiltonian of Eq.~\eqref{tdbg_ham} modifies to
\be 
h_{k,l} =  \hbar v_0 {\mathcal R}(\mp \theta/2)~[(\mathbb{I} + {\mathcal E}^T)  ] ({\bm k} - {\bm D}_{\xi}) \cdot (\xi \sigma_x, \sigma_y) + \frac{\delta}{2} ( \mathbb{I} \mp \sigma_z)~.
\ee

\noindent The complete derivation of strain-induced modifications is provided in 
Ref.~\cite{sinha_berry_2022_SI, Chakraborty2022_SI, he_NC2020_SI}. The strain matrix operates over the position of the twisted Dirac points given by
\be
{\bm D}_\xi = (\mathbb{I} - {\mathcal E}^{T}) {\bm K}^i_\xi - \xi {\bm A}~,
\ee
with ${\bm A}$ representing the gauge field that has the dimension of the reciprocal lattice vector. The appearance of the gauge field can be attributed to the fact that the strain causes the inter-atomic distance in each layer to become different in different directions. This results in the difference of hopping parameters, which displaces the Dirac point from its original position. The gauge potential ${\bm A}$ in terms of the strain matrix elements is given by
\be 
{\bm A} = \dfrac{\sqrt{3}}{{2a}} \beta ({\mathcal E}_{xx} - {\mathcal E}_{yy} , -2 {\mathcal E}_{xy})~.
\ee

\begin{figure*}[t]
    \centering
    \includegraphics[width=17.5cm]{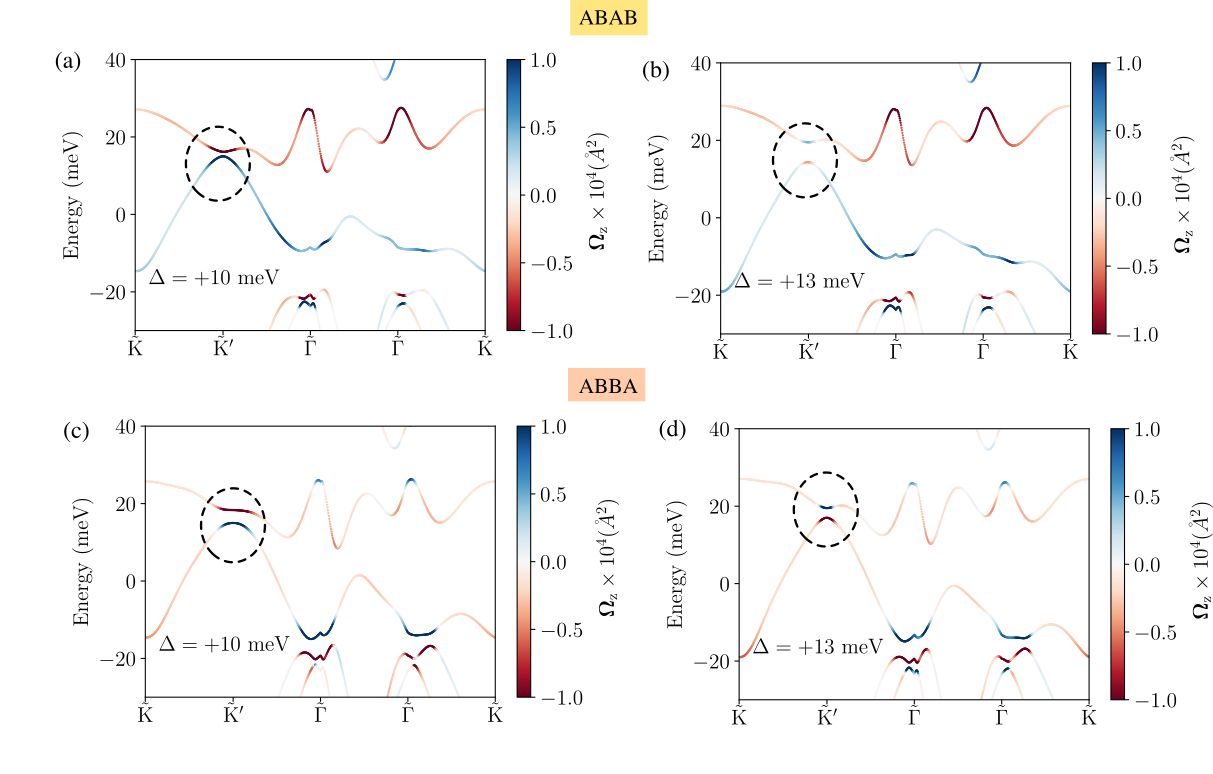}
    \caption{ \label{sfig:fig3_theory} \textbf{ Band dispersion around topological phase transition in presence of $\epsilon=0.2\%$ strain.} (a) and (b) represent the Berry curvature color-coded energy dispersion of AB-AB stacked TDBG for $\Delta=10$ meV and $\Delta=13$ meV respectively. The gap closing between CB and VB around $\Delta=11$ meV captures the topological transition. The Berry curvature changes sign across the two phases of phase transiton. Momentum dependent Berry curvature resolved band-dispersion of AB-BA for $\Delta=10$ meV and $\Delta=13$ meV are shown in (c) and (d). The sign change across topological transition is invariant with respect to stacking order. }
\end{figure*}
\noindent Here, $\beta=1.57$ and ${\mathcal E}_{ij}$ are the elements of the strain matrix. Strain also modifies the lattice vectors and consequently, the hopping matrices and the hopping vectors. We calculate the strained moir\'e vectors starting from un-rotated and un-strained lattice vectors. We obtain the lattice vectors for the strained lattice using ${\bm G}_{m}^{1, {\rm st}}=R_{-\frac{\theta}{2}} (1-{\mathcal E}^T){\bm b}_1 - R_{\frac{\theta}{2}} {\bm b}_1$ and ${\bm G}_{m}^{2, {\rm st}}=R_{-\frac{\theta}{2}} (1-{\mathcal E}^T){\bm b}_2 - R_{\frac{\theta}{2}} {\bm b}_2$~.

The band dispersions for the lowest energy valence and conduction bands in ABAB and ABBA stacking are shown in Fig.~\ref{sfig:fig1_theory} (a, b) and (c, d) for $\Delta=\pm 8$ eV in the presence of $0.2\%$ strain. Next, we calculate the Berry curvature (BC) which is defined as $\Omega^n_d= \frac{1}{2} \epsilon_{dac} \Omega_{ac}^n$, where
\be 
\Omega^n_{ac} = - 2 {\rm Im} \sum_{m \neq n  } \dfrac{ \langle  u_n |\partial_{k_a} {\mathcal H} | u_m \rangle \langle  u_m |\partial_{k_c} {\mathcal H} | u_n\rangle}{(\epsilon_n - \epsilon_m)^2} ~.
\ee

\noindent Here, $u_n$ is the periodic part of the Bloch wave-function where ${\mathcal H} | u_n \rangle = \epsilon_n | u_n \rangle $ corresponds to the moir\'{e} bands. In ABAB-stacked TDBG, the BC hotspot at the band edge reverses sign with the switching of electric field direction. In contrast, the sign of BC and the associated valley Chern number, $Z_2\equiv (C_K-C_{K^\prime})/2$ in ABBA-stacked TDBG remain unaffected by the electric field polarity. Here, $C_K$ and $C_{K^\prime}$ denote Chern numbers of the individual bands at $\mathbf{K}$ and $\mathbf{K}^\prime$ valleys, respectively. 

After analyzing the electronic band structure and the BC evolution in TDBG under varying electric fields, we now focus on the second-order non-linear  Hall response of the ABAB and ABBA stacked TDBG. The non-linear conductivity $\sigma_{abc}$ which links the non-linear current to the in-plane electric field, ${\bm E}$ via the relation $j^{(2)}_a = \sigma_{abc} E_b E_c$, where $a$, $b$ and $c$ represent spatial indices, is defined as 
\be \label{sigma_abc}
\sigma_{abc} = \epsilon_{abd} \dfrac{e^3 \tau}{\hbar^2} \Lambda_{dc}^\sigma~~~~  \mathrm{where,}~~\Lambda^\sigma_{dc} = - \sum_n \int [d{\bm k}]\dfrac{\partial f_0^n}{\partial k_c} \Omega^n_d~.
\ee
Here, $-e$ is the electronic charge, $\epsilon_{abd}$ is the anti-symmetric Levi-Civita tensor, $\tau$ is the scattering time and $\Lambda_{dc}^\sigma$ is the BCD. 

\begin{figure*}[t]
    \centering
    \includegraphics[width=16cm]{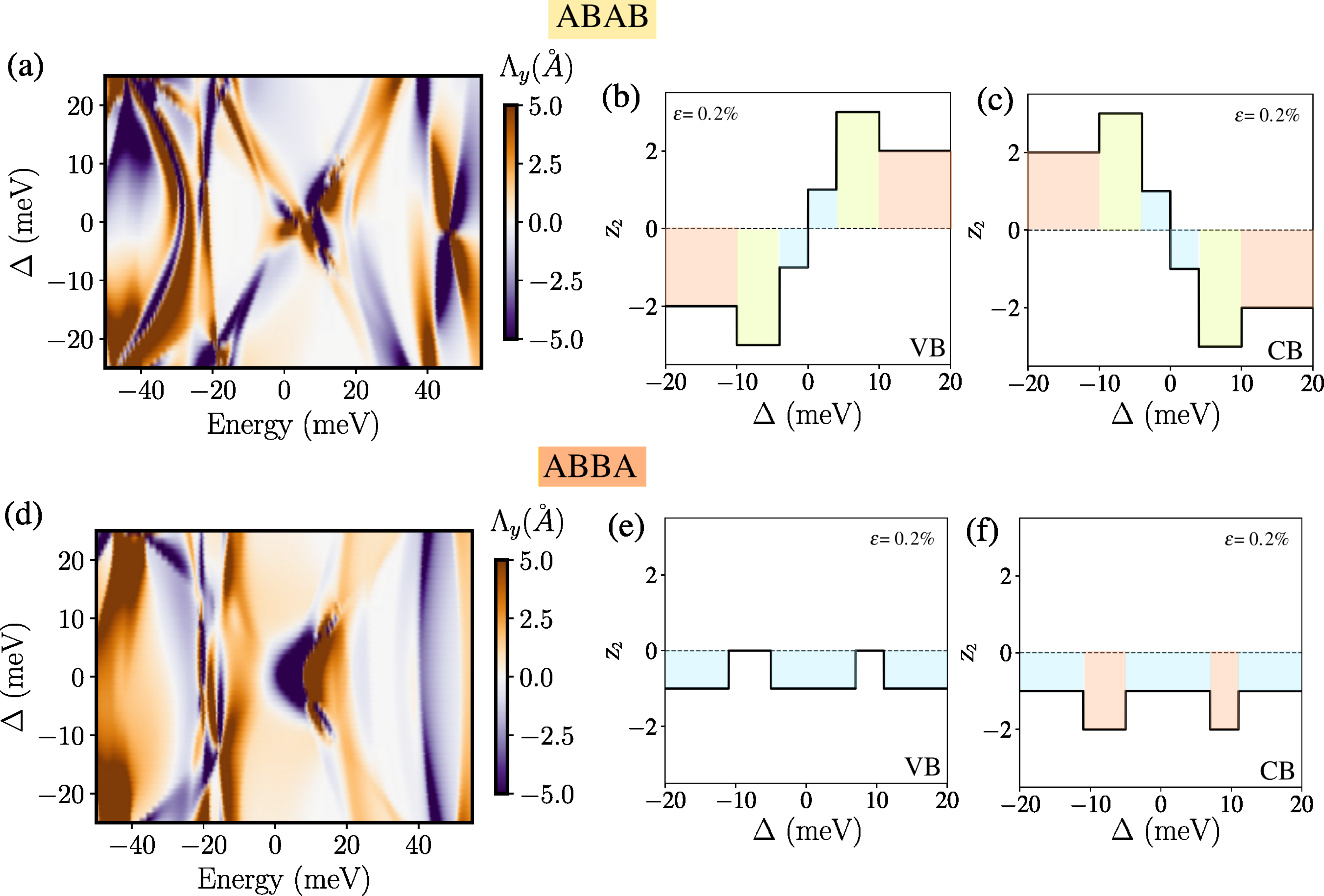}
    \caption{ \label{sfig:fig4_theory} \textbf{ BCD phase diagram and Chern number.} Variation of BCD over $\Delta-$Energy plane (a) for AB-AB, (d) for AB-BA in presence of strain $\epsilon=0.2\%$. The change in the valley chern number $Z_2$ of ABAB-TDBG for lowest valence band (VB) (b) and conduction band (CB) (c) for fixed strain $\epsilon=0.2\%$. The Chern number changes sign between positive and negative $\Delta$ values for both VB and CB. Similarly, the valley Chern numbers of VB and CB for $\epsilon=0.2\%$ are plotted in (e) and (f) for ABBA-TDBG. The valley Chern number is symmetric with respect to the change is the direction of electric field for ABBA-stacked TDBG. }
\end{figure*}
In Fig.~\ref{sfig:fig2_theory} (a,b) and (c,d) we plot the variation of $\lambda_y$ with energy near the charge neutrality for ABAB and ABBA stacking, considering two different values of the inter-layer potential $\Delta=$8 eV and 10 eV. Notably, for both the field strengths, reversing the electric field direction, as indicated by the sign of $\Delta$, results in a sign reversal of BCD in ABAB-stacked TDBG. However, for ABBA-stacked TDBG, the BCD sign remains unaffected by changes in the field direction.  The slight variation in the response magnitude between positive and negative $\Delta$ likely arises from the influence of strain, and other changes in the band structure.

Remarkably, the variation of the perpendicular electric field in both ABAB and ABBA-stacked TDBG can induce topological phase-transition of the valley-Chern type. Similar to the usual phase-transition in Chern insulators, the valley-Chern number changes in TDBG are also associated with the band gap closing at specific $k$ point. To illustrate this, we explicitly examined the evolution of the band movement in Fig.~\ref{sfig:fig3_theory} in presence of $\epsilon=0.2\%$ strain. The BC swaps between the band-touching edges between the consecutive bands across the transition at $\Delta=12$ meV as we see in Fig.~\ref{sfig:fig3_theory} (a,b) for ABAB and similarly in Fig.~\ref{sfig:fig3_theory} (c,d) for ABBA. We find multiple phase transitions in both ABAB- and ABBA-stacked TDBG on varying electric field. The distribution of $\lambda_y$ is plotted over the $\Delta-$Energy plane in Fig.~\ref{sfig:fig4_theory} (a, d). 

To track the changes in the band topology, we plot the valley Chern number $Z_2$ for ABAB stacked TDBG in Fig.~\ref{sfig:fig4_theory} (b), (c) for the lowest valence and conduction band for a fixed strain strength of $\epsilon=0.2\%$. We find three topological phase transitions within range $\Delta=$0 to 20 meV.  Interestingly, the valley Chern number switches sign with the sign change of $\Delta$. In contrast, the Chern histogram shown in Fig.~\ref{sfig:fig4_theory} (e, f) for ABBA VB and CB exhibit a symmetric nature with respect to positive and negative $\Delta$. This captures the distinct topological phases induced by the stacking order, which exhibit different behavior on reversing the polarity of the vertical electric field. We show below that this imprint of stacking order on band topology can be captured by the non-linear transport experiments.

While our theoretical model provides qualitative insights consistent with experimental conclusions, quantitative differences may arise due to the sensitivity of calculated BCD to strain direction (e.g., Fig. 4 of Pantaleon et al.~\cite{low_tunable_2021_SI}) and magnitude (Section V of Supplementary Information in Sinha et al.~\cite{sinha_berry_2022_SI}). 
We used a simplified uniaxial strain model along the zigzag direction of graphene, whereas an experimental strain is likely more complex and uncontrolled in magnitude or orientation. 
This can be one of the primary causes for the difference between the BCD values calculated in theory and obtained in experiments. 
For this reason, we do not focus on the magnitude of BCD, but use the BCD sign changes with the polarity of the perpendicular electric field to decipher the TDBG stacking order. 
Furthermore, the exact values of used parameters to calculate the band structure in twisted graphene systems are still being debated in the literature, even though they capture the essential physics. 
Despite these limitations, our theoretical framework effectively explains the contrasting BCD behavior observed in AB-AB and AB-BA stacked TDBG.

\section{Tight-Binding model of bilayer graphene}
To understand the stacking-mediated band-topology, in this section, we focus on the simple bilayer graphene (BLG) model to calculate the electronic bands, their BC and first moment, BCD.  In the following, we introduce  $4\times 4$ low energy tight binding (TB) model Hamiltonian for AA- and AB-stacked BLG model~\cite{Schaefer2021_SI,datta_nonlinear_2024_SI,JunaidIEEE_SI}. 

\subsection{Minimal TB model for AA-stacked BLG}
The low energy Hamiltonian for AA-stacked bilayer graphene (BLG) in the basis [$l^1_A,~l^1_B,~l^2_A,~l^2_B$] is expressed as,
\begin{center}\[
\begin{pmatrix}
    -\Delta & -\gamma_0\phi(\mathbf{k}) & \gamma_1 & 0 \\
    -\gamma_0\phi^*(\mathbf{k}) & -\Delta & 0 & \gamma_1 \\
    \gamma_1 & 0 & +\Delta & -\gamma_0 \phi(\mathbf{k}) \\
    0 & \gamma_1 & -\gamma_0\phi^*(\mathbf{k}) & +\Delta
\end{pmatrix}~,\]
\end{center}
where $l^i_\alpha$ represent the $\alpha$ sublattice of layer $i$. The geometric factor is given by $\phi(\mathbf{k})=\sum_{l=1}^3 e^{i\mathbf{k \cdot \delta_l}}$ where $\mathbf{\delta_l}$ denotes the positions of the three nearest B sublattice relative to the A sublattice or vice-versa within one monolayer graphene. The three connecting vectors can be expressed as $\mathbf{\delta_1}=(\mathbf{a}_1-\mathbf{a}_2)/3$, $\mathbf{\delta_2}=(\mathbf{a}_1+2\mathbf{a}_2)/3$, and $\mathbf{\delta_3}=-(2\mathbf{a}_1+\mathbf{a}_2)/3$. Here $\gamma_0$ and $\gamma_1$ are intra-layer and inter-layer hopping elements. We used the following model parameters $\gamma_0=3.16$ eV and $\gamma_1=0.381$~eV for our calculation. We disregard the negligible inter-layer hopping between non-dimeric sites in the model. The layer-dependent potential, $\Delta$, is used to account for the effect of the perpendicular electric field. 

The sole presence of an electric field alone is not sufficient to open a gap for AA-stacked BLG. Note that, the existence of a band gap is crucial for estimating the Berry curvature. To address this, we introduced a small artificial corrugation effect by incorporating a perturbative $H^{\mathrm{cor}}$ Hamiltonian to the original Hamiltonian. Here we set, $H^{\mathrm{cor}}_{11}=\delta_{cor}$, $H^{\mathrm{cor}}_{22}=H^{\mathrm{cor}}_{33}=H^{\mathrm{cor}}_{44}=-\delta_{cor}$~ and $H^{\mathrm{cor}}_{ij}=0$ for $i \neq j$ for the calculation of energy dispersion. $\delta_{cor}$ represent the artifical corrugation gap. The energy dispersion of the artificially corrugated AA-stacked BLG with $\delta_{cor}=$0.08 eV for $\Delta=\pm 0.1$ eV is included in Fig.~5 of the main manuscript.

\subsection{Minimal TB model for AB-stacked BLG}
In this section, we present the low energy minimal Hamiltonian for AB-stacked namely Bernal BLG in the  [$l^1_A,~l^1_B,~l^2_A,~l^2_B$] basis. The Hamiltonian, accounting for the effect of the electric field, has the following expression, 
\begin{center}
\[
\begin{pmatrix}
    -\Delta & -\gamma_0\phi(\mathbf{k}) & 0 & 0 \\
    -\gamma_0\phi^*(\mathbf{k}) & -\Delta & \gamma_1 & 0 \\
    0 & \gamma_1 & +\Delta & -\gamma_0\phi(\mathbf{k}) \\
    0 & 0 & -\gamma_0\phi^*(\mathbf{k}) & +\Delta
\end{pmatrix}~,
\]
\end{center}   
The incorporation of the perpendicular electric field $\Delta$ opens up a gap between the valence and conduction band, breaking the inversion symmetry of the AB-BLG. The energy dispersion of the pristine AB-stacked BLG for $\Delta=\pm 0.1$ eV are included in Fig.~5 of the main manuscript.

\subsection{Impact of Strain and generation of BCD}

The breakdown of inversion symmetry is essential to have a finite Berry curvature in systems preserving time-reversal symmetry. However, pristine BLG has three-fold rotational symmetry (C3) that forbids BCD. Therefore, we apply a uniaxial strain in the graphene disrupting the C3 symmetry to overcome this limitation. Consider a uniaxial strain of strength $\epsilon$, applied along an arbitrary angle $\phi$ relative to the zig-zag direction. It can be described by the following strain tensor 
\be \label{strain}
{\mathcal E} = \epsilon
\begin{pmatrix}
-\cos^2 \phi + \nu \sin^2 \phi & -(1+ \nu) \sin \phi \cos \phi \\
-(1+ \nu) \sin \phi \cos \phi & -\sin^2 \phi + \nu \cos^2 \phi  
\end{pmatrix},
\ee

\begin{figure*}[t]
    \centering
    \includegraphics[width=15cm]{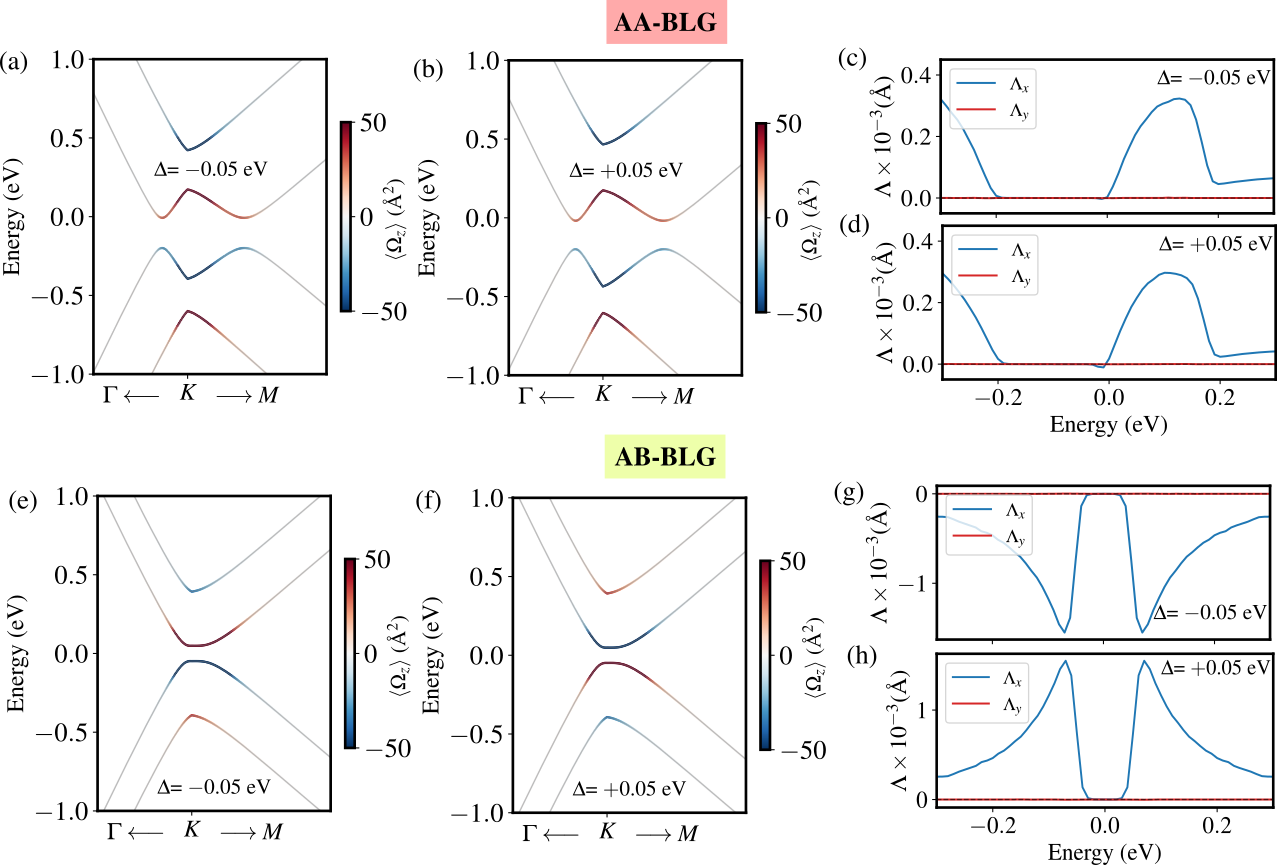}
    \caption{ \label{sfig:fig5_theory} \textbf{}
    \textbf{Band dispersion and Berry curvature dipole of strained BLG.} (a) and (b) show the strained energy dispersion of artificial corrugated AA-stacked BLG for $\Delta=-0.05$ eV and $0.05$ eV respectively. We choose strain $\epsilon=1.0\%$ for our plot. The corresponding BCD over the energy for $\Delta=\mp0.05$ are plotted in (c) and (d). The strained band-structures around high-symmetry $\mathbf{K}$ point are shown in (e) and (f) for $\Delta=\mp0.05$. The corresponding BCD of AB-stacked BLG over the energy for $\Delta=\mp0.05$ are plotted in (g) and (h). The BCD of AB-BLG changes sign with direction flip of the electric field. }
\end{figure*}

\noindent Here, $\epsilon$ is the strength of the strain, $\nu$ is the Poisson ratio ($\sim 0.16$ for graphene) and $\phi$ is the strain angle w.r.t zigzag direction of graphene.  The applied uniaxial strain distorts the lattice structure with $\mathbf{a} \rightarrow \mathbf{a}^\prime =(\mathbb{1}+\mathcal{E}) \mathbf{a}$ and renormalizes reciprocal lattice vectors with $\mathbf{b} \rightarrow \mathbf{b}^\prime = (\mathbb{1}-\mathcal{E}^T) \mathbf{b}$. The corresponding hopping parameters modifies to $\gamma_0 \rightarrow \gamma_0 e^{-\beta(\frac{|\delta^\prime|}{|a_0|} -1)}$, where $\beta= 1.57$. Here $\mathbf{a_0}\equiv \frac{\mathbf{a}}{\sqrt{3}}$ is the bond length between adjacent carbon atoms of monolayer graphene, and we choose $\phi=0$, i.e. uniaxial strain along the zigzag direction of the graphene layer.

The energy dispersions of the AA and AB stacked BLG in the presence of $\epsilon=1\%$ strain are plotted in Fig.~\ref{sfig:fig5_theory}. We choose $\Delta=-0.05$ eV along with the artificial corrugation potential $\delta_{cor}=0.2$ eV to calculate the band dispersion of AA stacked BLG as shown in Fig.~\ref{sfig:fig5_theory} (a). The energy dispersion around high symmetry $\mathbf{K}-$point for AB stacked BLG is shown in Fig.~\ref{sfig:fig5_theory} (c). Clearly, the BC hot spot lies at the band edge. For both stacking, the lowest conduction and valence band pairs host opposite BC.  Next, we change the sign of the $\Delta$ to account the change of direction of perpendicular electric field and plot the band dispersion in Fig.~\ref{sfig:fig5_theory} (b) and (f) for AA- and AB-stacked BLG. The switching of the electric field, changes the sign of BC for AB-BLG whereas the nature of BC remains same for AA-BLG.

To investigate the variation of BCD with change of electric field direction we plot $\Lambda_x$ and $\Lambda_y$ for $\Delta=\mp 0.05$ for AA-stacked corrugated BLG (see Fig.~\ref{sfig:fig5_theory} c, d) and Bernal AB-stacked BLG (see Fig.~\ref{sfig:fig5_theory} g, h). The BCD of AA-stacked BLG remains unaffected by the sign change of the electric field. However, the BCD changes sign with electric field direction for AB-stacked BLG. Hence, the BCD variation on reversing the direction of the electric field is sensitive to the stacking order.

\subsection{Variation of Berry curvature at band edge with perpendicular electric field}

\begin{figure*}[h]
    \centering
    \includegraphics[width=14cm]{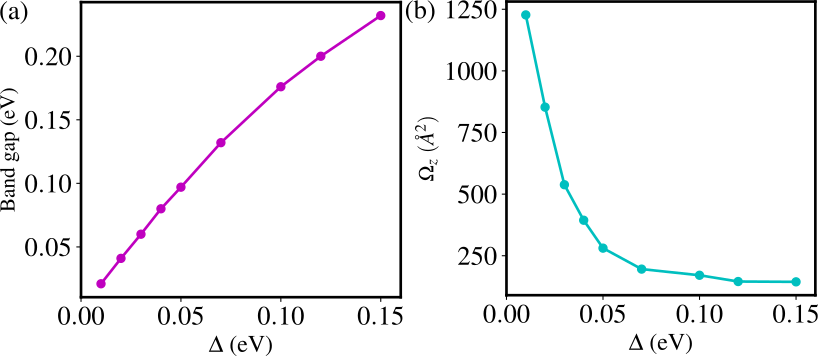}
    \caption{ \label{sfig:fig6_theory} \textbf{}
    \textbf{Variation of band-gap and BC with electric field of BLG.} (a) The magnitude of direct band gap between the lowest valence and conduction band of AB-stacked BLG for different values of $\Delta$. (b) shows the value of BC at valence band edge with variation of $\Delta$. The BC falls sharply for low $\Delta$ values. For the higher value of perpendicular electric field, the rate of change of BC is relatively small.
    }
\end{figure*}

In this section, we explore the change in the magnitude of the BC hot-spot with perpendicular electric field. We systematically increased the $\Delta$ from $0.01$ eV to $0.15$ eV for Bernal AB-stacked BLG. The change in the direct band-gap at the charge-neutrality with variation of $\Delta$ is plotted in Fig.~\ref{sfig:fig6_theory} (a). The band-gap increases almost linearly with the increment of perpendicular electric field $\Delta$. We plot the magnitude of BC at valence band-edge for the corresponding $\Delta$ range. Variation of $\Omega_z$ at VB w.r.t $\Delta$ is shown in Fig.~\ref{sfig:fig6_theory} (b).  Interestingly, the BC changes sharply for the low $\Delta$ values. The rate of change in BC decreases for the larger values of $\Delta$ which is consistent with the findings of Ref.~\cite{Shimazaki2015_SI}. For this reason, we choose high $\Delta$ range for the fitting of the experimental data. \\

\section{Device Fabrication}
ABAB-TDBG and ABBA-TDBG samples were fabricated using the ‘cut-and-stack’ method. Bilayer graphene and h-BN flakes (thickness: 20-40~nm) were exfoliated onto a 285~nm thick SiO$_2$/Si++ substrates. A Bernal-stacked~(AB) bilayer graphene flake was first cut into two pieces using a tapered optical fiber scalpel~\cite{varma_sangani_facile_2020_SI} to ensure that both pieces shared the same crystallographic axis.

\begin{figure*}[h]
    \centering
    \includegraphics[width=14cm]{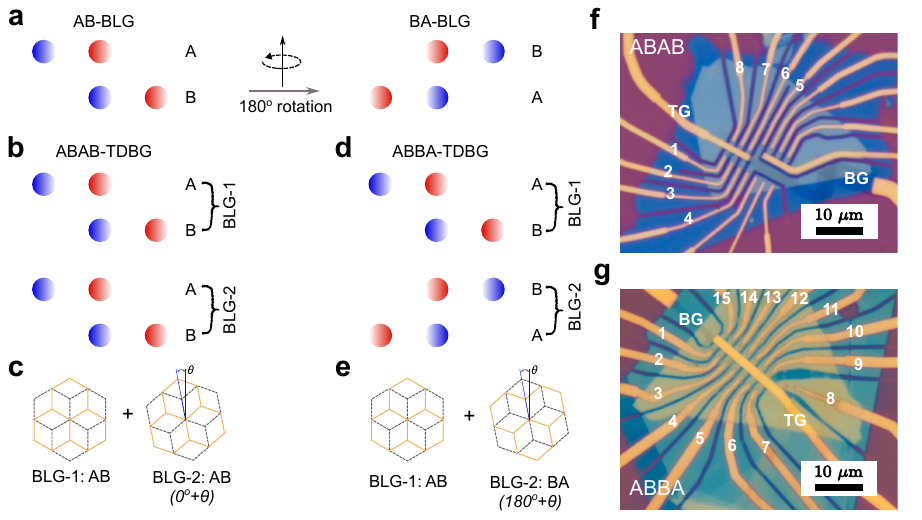}
    \caption{ \label{sfig:fab} \textbf{}
    \textbf{Fabrication of AB-AB and AB-BA TDBG devices.} 
    (a) Atomic arrangements in AB-stacked bilayer graphene (AB-BLG) and its transformation into BA-stacked bilayer graphene (BA-BLG) upon a 180$^\circ$ rotation. Red and blue spheres represent the two distinct sublattices of graphene. 
    (b) Two AB-stacked bilayer graphene (BLG-1 and BLG-2) layers are stacked on top of each other, forming the AB-AB TDBG. 
    (c) Two AB-BLGs are twisted relative to each other by an angle $\theta$, resulting in the twisted AB-AB TDBG structure. The dashed and solid lines indicate the top and bottom layers within a single BLG. 
    (d) The AB-stacked BLG-1 and BA-stacked BLG-2, created by an additional 180$^\circ$ rotation of AB-BLG, are stacked together to form AB-BA TDBG. 
    (e) Two BLGs are twisted relative to each other by 180$^\circ+\theta$, forming the twisted AB-BA TDBG structure. 
    (f) Optical image of the fabricated AB-AB TDBG device with a relative twist angle of 1.43$^\circ$. Top and bottom gates (TG and BG) and numbered contact electrodes are visible. The width and the length of the region used for measurements are 1~$\mathrm{\mu m}$ and 1.75~$\mathrm{\mu m}$ respectively.
    (g) Optical image of the fabricated AB-BA TDBG device with a relative twist angle of 1.4$^\circ$. Top and bottom gates (TG and BG) and numbered contact electrodes are labeled. Scale bars represent 10~$\mathrm{\mu m}$. The width and the length of the region used for measurements are 1.5~$\mathrm{\mu m}$ and 1.95~$\mathrm{\mu m}$ respectively.
    }
\end{figure*}

To assemble the heterostructures, an h-BN flake, serving as the dielectric layer for the top gate, was picked up using a PC (Poly (Bisphenol A carbonate)) + PDMS (polydimethylsiloxane) stamp at 110~$^\circ$C. One half of the bilayer graphene was then picked up at 90~$^\circ$C using this h-BN flake. The second half was subsequently picked up with a rotation angle of $\theta$ (180$^\circ$+$\theta$) relative to the first half, forming either the ABAB-TDBG (ABBA-TDBG) configuration. The rationale for the 180$^\circ$ rotation is schematically depicted in Fig.~\ref{sfig:fab}. Following this, an additional h-BN flake, serving as the dielectric layer for the bottom gate, and a few-layer graphite flake were sequentially picked up at 90-100~$^\circ$C.

The stack was transferred onto Si++/SiO$_2$ substrates that had been treated with O$_2$ reactive ion etching (40 sccm O$_2$, power: 25 W, pressure: 1 Pa) to prepare the surface. Residual PC was removed by rinsing the samples in chloroform. The heterostructures were then patterned into dual-gate Hall bar devices using electron beam lithography. Cr/Au (5 nm/60 nm) was deposited to form the top gate electrode, while a few-layer graphite was used as the back-gate electrode.

To establish edge contacts, the top h-BN layer was etched using CHF$_3$/O$_2$ plasma (CHF$_3$: 40 sccm, O$_2$: 4 sccm, power: 60 W, pressure: 1 Pa). 
Finally, Cr/Pd/Au (5 nm/20 nm/40 nm) contacts were deposited using electron beam evaporation, following in situ Ar plasma cleaning.
A non-zero fixed global gate voltage was applied to the heavily doped silicon to further reduce the contact resistances.
Ohmic behavior of the two probe contact resistances were verified. 

\section{Estimation of twist angle of AB-AB and AB-BA TDBG devices}
\subsection{Device architecture and twist angle estimation}
We performed low-temperature transport measurements primarily at 1.2 K-1.5 K, unless otherwise specified, for the AB-AB and AB-BA TDBG devices, using a $\mathrm{He_4}$ flow cryostat.
A current of 100 nA with a frequency ($\omega$) of 177 Hz was applied and the four-probe longitudinal resistance R$_{xx}=$V$_{xx}^\omega/I_\omega$ was measured using SR-830 lock-in amplifier, following amplification using a DL instrument voltage preamplifier.

The charge density ($n$) and the perpendicular electric displacement field ($D$) were determined using the formulas $n = (C_\text{BG}V_\text{BG} + C_\text{TG}V_\text{TG})/e - n_0$ and $D = (C_\text{BG}V_\text{BG} - C_\text{TG}V_\text{TG})/2$.
Here, $C_\text{TG}$ and $C_\text{BG}$ represent the capacitance per unit area of the top and back gates, respectively, $e$ denotes the electron charge, and $n_0$ is the offset in charge density due to unintentional doping.
$V_\text{BG}$ and $V_\text{TG}$ denote the DC voltage applied to the back gate and top gate, respectively.
The capacitance values were derived using the hBN dielectric thickness of a gate and the slope of R$_{xx}$ peak at the charge neutrality in the $V_\text{BG}$-$V_\text{TG}$ plane.
The values were subsequently verified via analysis of magneto-transport features, such as the positions of Brown-Zak oscillations (see Fig.~\ref{fig:SI_twist_angle_ABAB}b) and the tracing of Landau levels (see Fig.~\ref{fig:SI_twist_angle_ABAB}b and Fig.~\ref{fig:SI_twist_angle_ABBA}b) from the fan diagram.

Figure~\ref{fig:SI_twist_angle_ABAB}a shows the measured R$_{xx}$ in the full parameter space of $n$ and perpendicular electric field \D{}.
The fan diagrams at $D/\epsilon_0 = 0$ for AB-AB (AB-BA), displayed in Fig.~\ref{fig:SI_twist_angle_ABAB}b (Fig.~\ref{fig:SI_twist_angle_ABBA}b), were obtained with magnetic fields reaching up to 13.6 T.
The overlaid lines on the fan diagram, originating from $\nu = 0$ and $\pm 4$, correspond to different Landau levels.
The horizontal grey dashed lines in Fig.~\ref{fig:SI_twist_angle_ABAB}b correspond to Brown-Zak oscillations, resulting in R$_{xx}$ dips visible most clearly on the (-12,0) line for the AB-AB device.

To determine the twist angle $(\theta)$, we estimated the value of $n_S$ (see Fig.~\ref{fig:SI_twist_angle_ABAB}c for AB-AB device-1 and Fig.~\ref{fig:SI_twist_angle_ABBA}a for AB-BA device used in the main manuscript) from our low-temperature electron transport measurements.
The twist angle is then extracted utilizing the relation $n_S=8\theta^2/\sqrt{3}a^2$.
In this equation, $n_S$ denotes the charge carrier density corresponding to the full filling of the moir\'e band $(\nu = \pm 4)$, and $a = 0.246\ \mathrm{nm}$ is the lattice constant of graphene.
The AB-AB TDBG device-2 with an angle of 1.1$^\circ$, used in Section~\ref{sec:ABABdev2}, has been characterized similarly~\cite{adak_perpendicular_2022_SI}.

\subsection{Twist angle inhomogeneity and evidence of strain}
\begin{figure*}[h]
    \centering
    \includegraphics[width=16cm]{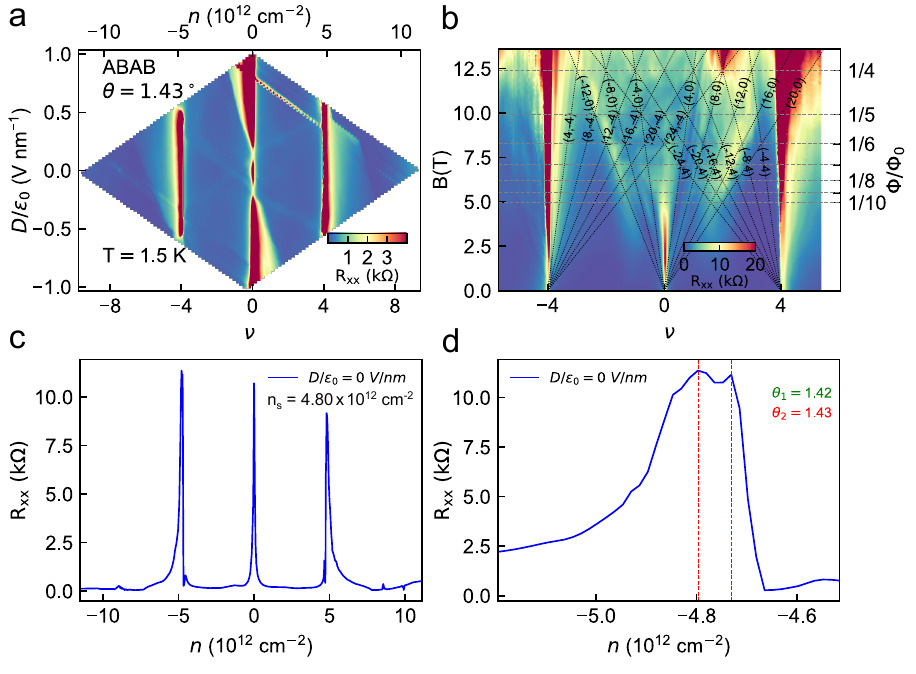}
    \caption{ \label{fig:SI_twist_angle_ABAB} \textbf{Estimation of twist angle and signature of strain in AB-AB TDBG device-1.}
    \textbf{a,}~Four-probe longitudinal resistance \Rxx{} of AB-AB TDBG device-1 (shown in Fig.~2a of the main manuscript) as a function of filling factor ($\nu$) and perpendicular electric field (\D) for extended $\nu$ and \D{} range.
    The top axis indicates the charge density ($n$).
    \textbf{b,}~Colorscale plot of \Rxx{} as a function of $\nu$ and perpendicular magnetic field $B$.
    The ($\nu_\text{LL}$, $\nu$) values indicate the Landau level filling factor $\nu_\text{LL}$ that originates from the filling $\nu$ at $B=0$, and the corresponding dotted lines coincide with local resistance minima.
    The right axis shows the corresponding values of $\phi/\phi_0$, where $\phi$ is the magnetic flux through a \moire unit cell and $\phi_0$ is the magnetic flux quantum.
    The horizontal dashed lines correspond to a decrease in \Rxx{} due to Brown-Zak oscillations that occur at simple fractions of $\phi/\phi_0$.
    \textbf{c, d}~\Rxx{} vs $n$ lineslice at \D$=0$~\Vbynm{} for the full $n$ range measured (c), and zoomed in close to $n=-n_S$ (d).
    The two subpeaks correspond to two slightly different twist angles due to strain in the sample.
    }
\end{figure*}

\begin{figure*}[h]
    \centering
    \includegraphics[width=16cm]{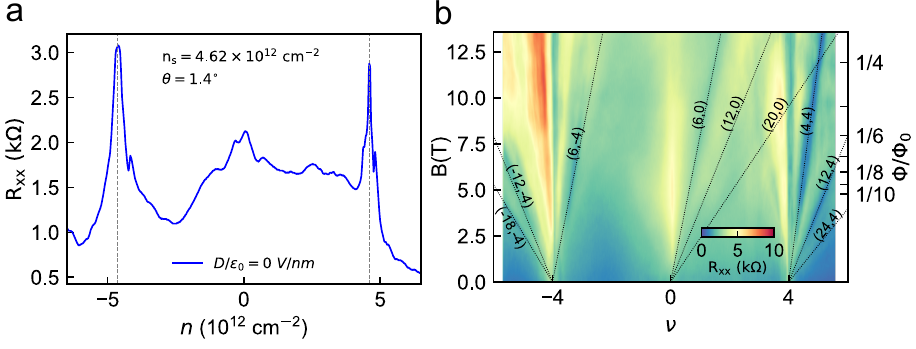}
    \caption{ \label{fig:SI_twist_angle_ABBA} \textbf{Estimation of twist angle in AB-BA TDBG device.}
    \textbf{a,}~\Rxx{} vs $n$ lineslice at \D$=0$~\Vbynm{} for the full $n$ range measured at $T=1.5$~K.
    The extracted value of $n_S$ indicates a twist angle of $\theta=1.4^\circ$.
    \textbf{b,}~Colorscale plot of \Rxx{} as a function of $\nu$ and perpendicular magnetic field $B$, measured at $T=300 mK$.
    The ($\nu_\text{LL}$, $\nu$) values indicate the Landau level filling factor $\nu_\text{LL}$ that originates from the filling $\nu$ at $B=0$, and the corresponding dotted lines coincide with a resistance minima.
    The right axis shows the corresponding values of $\phi/\phi_0$, where $\phi$ is the magnetic flux through a \moire unit cell and $\phi_0$ is the magnetic flux quantum.
    }
\end{figure*}

Strain is a crucial parameter in twisted graphene devices, as it breaks the $C_3$ symmetry~\cite{kazmierczak_strain_2021_SI,mcgilly_visualization_2020_SI} and induces a finite Berry curvature dipole.
Evidence of strain in our TDBG devices is demonstrated by the presence of angle inhomogeneity.
Specifically, if two different predominant angles exist within our device, they correspond to two distinct peaks in R$_\text{xx}$ vs. $n$ dependence at full-filling ($n=\pm n_S$).
Our transport measurements revealed two sub-peaks at the moir\'e peak (see Fig.~\ref{fig:SI_twist_angle_ABAB}d), indicating two different $n_S$ values.
From this observation, we estimated $n_{S1} = 4.73\times10^{12}\ \mathrm{cm^{-2}}$ and $n_{S2} =4.80\times10^{12}\ \mathrm{cm^{-2}}$, which correspond to twist angles $\theta_1 = 1.42^{\circ}$ and $\theta_2 = 1.43^{\circ}$, respectively, in the AB-AB TDBG device-1 used in the main manuscript. 
Given that $\Delta \lambda/\lambda = \Delta \theta/\theta,$ $\lambda$ being the \moire wavelength, the strain in our AB-AB device is estimated to be, $(\theta_2-\theta_1) / \theta_{avg}=.007$, where $\theta_{avg}=(\theta_1+\theta_2)/2$.

\section{Estimation of band gap for 1.43$^\circ$ AB-AB TDBG device}
\begin{figure*}[h]
    \centering
    \includegraphics[width=16cm]{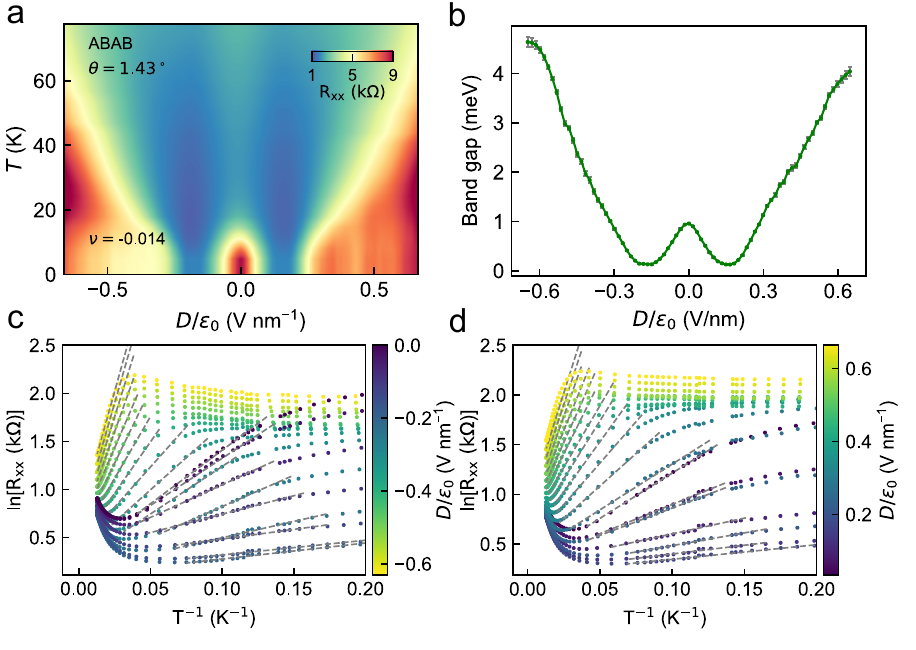}
    \caption{ \label{fig:SI_Bandgap_ABAB} \textbf{Bandgap estimation near charge neutrality point for AB-AB TDBG device-1.}
    \textbf{a,}~Longitudinal resistance \Rxx{} as a function of \D{} and temperature $T$.
    \textbf{b,}~The extracted band gap as a function of \D.
    The vertical gray bars represent the errors in bandgap extraction.
    \textbf{c, d,}~Arrhenius band gap extraction by fitting the linear regime (dashed gray lines) in ln (\Rxx) vs 1/$T$ for $D<0$ (c) and $D>0$ (d).
    The color of data points indicates the corresponding value of \D{} for a particular ln (\Rxx) vs 1/$T$ plot.
    }
\end{figure*}

We employed Arrhenius fitting to estimate the band gap of AB-AB device-1 and study its evolution under a perpendicular electric field at the CNP. 
$R_{xx}$ vs. \D{} dependence is measured by varying temperature upto $\approx$80~K (Fig.~\ref{fig:SI_Bandgap_ABAB}a).
A linear fitting of the natural logarithm ln($R_{xx}$) with the inverse of the temperature ($T^{-1}$) is done, where the slope is directly related to the band gap.
Figure \ref{fig:SI_Bandgap_ABAB}c (Figure \ref{fig:SI_Bandgap_ABAB}d) shows the linear fit in grey dashed lines for different negative (positive) values of \D{}.
The Arrhenius equation, $R_{xx}(T) \propto e^{E_g / 2k_B T}$, where $E_g$ is the band gap and $k_B$ is the Boltzmann constant, allows us to extract the band gap from the temperature-dependent resistance data.

Our observations reveal that the band gap of 1.43$^\circ$ AB-AB TDBG evolves with the applied perpendicular electric field at the CNP (Fig.~\ref{fig:SI_Bandgap_ABAB}b). 
As we varied the displacement field, we noticed the band gap initially decreased, leading to a band touching, and then reopened at a displacement field of approximately \D$ \approx \pm 0.17\ \mathrm{V\ nm^{-1}}$.

\section{Additional characterization of nonlinear voltage in AB-AB and AB-BA TDBG}
\begin{figure*}[h]
    \centering
    \includegraphics[width=16cm]{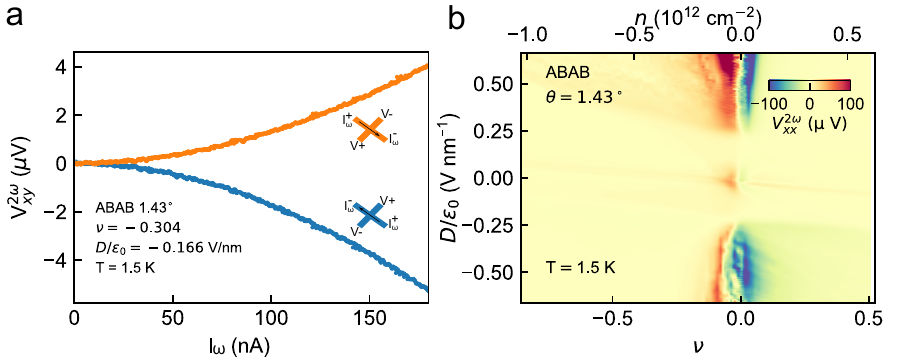}
    \caption{ \label{fig:SI_characterise_ABAB} \textbf{Further characterisation of nonlinear voltage in AB-AB stacked TDBG device-1.}
    \textbf{a,}~Quadratic scaling of \Vxytw{} with channel current I$_\mathrm{\omega}$ at a fixed $\nu$ and \D.
    The \Vxytw{} flips sign when the orientation of both the current and the voltage probes are flipped together.
    This indicates the second-order nature of the measured \Vxytw.
    \textbf{b,}~Colorscale plot of longitudinal nonlinear voltage \Vxxtw{} as a function of $\nu$ and \D{} close to the charge neutrality gap.
    The top axis shows the corresponding charge density $n$.
    }
\end{figure*}

\begin{figure*}[h]
    \centering
    \includegraphics[width=16cm]{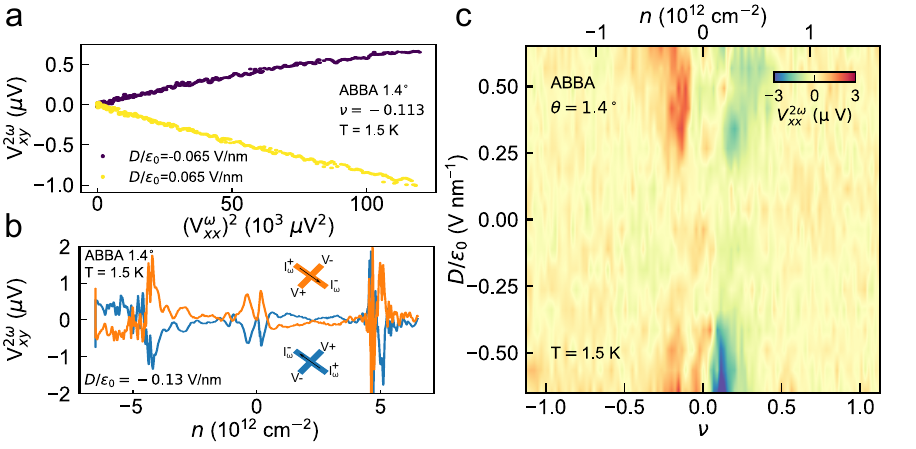}
    \caption{ \label{fig:SI_characterise_ABBA} \textbf{Further characterization of nonlinear voltage in AB-BA stacked TDBG device.}
    \textbf{a,}~Variation of \Vxytw{} with \Vxxwsq{} for a fixed filling factor $\nu=-0.113$, for two different polarities of the perpendicular electric field \D{}. 
    A linear behavior of V$_{xy}^{2\omega}$ with (V$_{xx}^{\omega}$)$^2$ verifies the quadratic dependence of V$_{xy}^{2\omega}$ on current.
    \textbf{b,}~\Vxytw{} vs $n$ lineslice for a fixed \D{}, for two different configuration of current and voltage probes.
    The \Vxytw{} flips sign when the orientation of both the current and the voltage measuring probes are flipped together.
    This indicates the second-order nature of the measured \Vxytw.
    \textbf{c,}~Colorscale plot of longitudinal nonlinear voltage \Vxxtw{} as a function of $\nu$ and \D{} close to the charge neutrality gap.
    The top axis shows the corresponding charge density $n$.
    }
\end{figure*}

The nonlinear Hall voltage $V_{xy}^{2\omega}$ exhibits a quadratic dependence on the applied current $(I_{\omega})$ having frequency $\omega$. 
This quadratic dependence arises because the nonlinear Hall effect is a second-order response to the external in-plane electric field $(E_{xx}^{\omega})$, making it sensitive to both the magnitude and direction of the applied current. 
Specifically, the nonlinear Hall voltage can be described by $V_{xy}^{2\omega} \propto (E_{xx}^{\omega})^2 \propto (I_{\omega})^2$, indicating that as the current increases, the nonlinear Hall voltage grows quadratically. 
The linear longitudinal voltage $V_{xx}^{\omega}$ is directly proportional to the current, following Ohm’s law, $V_{xx}^{\omega} = R_{xx} I_{\omega}$, where $R_{xx}$ is the longitudinal resistance. 
Hence, as the current is increased, $V_{xy}^{2\omega}$ changes linearly with $(V_{xx}^{\omega})^2$.
In Fig.~\ref{fig:SI_characterise_ABAB}a, we have demonstrated the quadratic dependence of the nonlinear Hall voltage on current up to 190 nA for the $1.43^{\circ}$ AB-AB device. 
Figure~\ref{fig:SI_characterise_ABBA}a illustrates the linear dependence of $V_{xy}^{2\omega}$ on $(V_{xx}^{\omega})^2$ for the $1.40^{\circ}$ AB-BA device.

Another characteristic feature of the nonlinear Hall effect is that when both the direction of current and voltage probes are reversed, $V_{xy}^{2\omega}$ picks up a negative sign. 
This behavior is depicted in Fig.~\ref{fig:SI_characterise_ABAB}a for the AB-AB device with $I_{\omega}$ on the x-axis, and in Fig.~\ref{fig:SI_characterise_ABBA}b with $n$ on the x-axis for the AB-BA device.

In addition to the nonlinear Hall voltage ($V_{xy}^{2\omega}$), we also observed a nonlinear longitudinal voltage ($V_{xx}^{2\omega}$) in our TDBG (Fig.~\ref{fig:SI_characterise_ABAB}b for AB-AB and Fig.~\ref{fig:SI_characterise_ABBA}c for AB-BA TDBG) having twist angle of $\approx$1.4$^\circ$.
The source of the nonlinear longitudinal voltage is primarily attributed to extrinsic scattering mechanisms~\cite{duan_giant_2022_SI,he_graphene_2022_SI,datta_nonlinear_2024_SI}. 
These include skew-scattering and side-jump scattering processes, which are influenced by disorder potentials in the system.
A recent proposal also suggests an intrinsic mechanism that generates a nonzero dissipative longitudinal second-order voltage, governed by the Berry connection polarizability induced by quantum metric~\cite{das_intrinsic_2023_prb_SI}.
Here we note that the scaling analysis (discussed in the next section) that we use to extract the BCD by parametrically varying \D{}, is applicable to our devices with non-zero skew-scattering.

\section{Scaling Analysis and local intercept analysis}
\subsection{Scaling Analysis}
The nonlinear Hall voltage \Vxytw{} can originate from both the intrinsic (Berry curvature dipole) and the extrinsic mechanisms (such as side-jump scattering, skew-scattering).
This is similar to the linear anomalous Hall voltage that can originate from both the intrinsic (Berry curvature) and the extrinsic scattering mechanisms~\cite{nagaosa_anomalous_2010-1_SI}.
Recent developments in the field suggest using a scaling relation to extract the Berry curvature dipole governed intrinsic contribution to \Vxytw{}.

Accounting for all the mechanisms, a general scaling relation for the experimentally measured normalized NLH signal \Vxytwnorm{}, can be written as,~\cite{du_disorder-induced_2019_SI} 
\be \label{eq:normNLHth}
\Vxytwnormeq = {\mathcal C}^{in} + \sum_i {\mathcal C}_i^{sj} \dfrac{\rho_i}{\rho_{xx}} + \sum_{i,j} {\mathcal C}_{ij}^{sk1} \dfrac{\rho_i \rho_j}{\rho_{xx}^2} + \sum_i {\mathcal C}_i^{sk2} \dfrac{\rho_i}{\rho_{xx}^2} ~.
\ee
Here, $i,j$ represent different sources of scattering.
The coefficients stand for Berry curvature dipole ($\mathcal{C}^{in}$), side-jump ($\mathcal{C}^{sj}$), and skew-scattering ($\mathcal{C}^{sk}$) contributions ({\it sk1} and {\it sk2} represent two different kinds of skew scattering).
Considering only two sources of scattering, the static (impurities) and dynamic (phonon), we can express the above equation as
\be  \label{sc}
\Vxytwnormeq=\dfrac{1}{\rho_{xx}^2}\left({\mathcal C}_1 \rho_{xx0} +{\mathcal C}_2 \rho_{xx0}^2+  {\mathcal C}_3 \rho_{xx0} \rho_{xxT} +   {\mathcal C}_4 \rho_{xxT}^2 \right)~.
\ee
Here, $\rho_{xx0}$ is the zero temperature residual resistivity due to static impurities, and $\rho_{xxT}=\rho_{xx}-\rho_{xx0}$ is the contribution from phonons at finite temperature.
The new parameter set in Eq.~\eqref{sc} can be obtained from the old one as
\bea
{\mathcal C}_1&=&{\mathcal C}^{sk2}_{0} ;~~{\mathcal C}_2={\mathcal C}^{in} +{\mathcal C}^{sj}_0+ {\mathcal C}^{sk1}_{00}, \\
{\mathcal C}_3&=&2{\mathcal C}^{in} +{\mathcal C}^{sj}_0+{\mathcal C}^{sj}_1+ {\mathcal C}^{sk1}_{01}, \\
{\mathcal C}_4&=&{\mathcal C}^{in} +{\mathcal C}^{sj}_1+ {\mathcal C}^{sk1}_{11} .
\eea
Here, the indices 0 and 1 stand for static and phonon scattering sources, respectively.
The coefficients $\mathcal{C}_{00}^{sk1}$, $\mathcal{C}_{11}^{sk1}$, $\mathcal{C}_{01}^{sk1}$ represent the skew-scattering contributions of the cross terms $\rho_i\rho_j/\rho_{xx}^2$ (where $i, j=0, 1$) to the normalized NLH voltage in Eq.~\ref{eq:normNLHth}.
$\rho_i$ represents the resistivity contribution of $i$-th type of scattering source to the longitudinal resistivity $\rho_{xx}=\Sigma_i \rho_i$. 
We can rewrite the scaling relation in terms of the conductivity $\sigma_{xx}$ as
\be \label{ratio_2}
\Vxytwnormeq -{\mathcal C}_1 \sigma_{xx0}^{-1} \sigma_{xx}^2 =\left({\mathcal C}_2 + {\mathcal C}_4 - {\mathcal C}_3 \right) \sigma_{xx0}^{-2} \sigma_{xx}^2
+ \left( {\mathcal C}_3 - 2{\mathcal C}_4\right) \sigma_{xx0}^{-1} \sigma_{xx}+   {\mathcal C}_4~.
\ee
Here, $\sigma_{xx0}$ is the residual conductivity.
After a rearrangement of the terms, Eq.~\ref{ratio_2} can be expressed as
\bea \label{ratio_3}
\Vxytwnormeq -{\mathcal C}_1 \sigma_{xx0}^{-1} \sigma_{xx}^2 =\left({\mathcal C}_2 - {\mathcal C}_4 \right) \sigma_{xx0}^{-2} \sigma_{xx}^2
+ \left( {\mathcal C}_3 - 2{\mathcal C}_4\right) \left(\sigma_{xx0}^{-1} \sigma_{xx}-\sigma_{xx0}^{-2} \sigma_{xx}^2 \right)+   {\mathcal C}_4~.
\eea
In the low-temperature limit in which we have done our experiment, the phonon contribution to the conductivity can be assumed to be small, and hence we consider $\sigma_{xx}\approx\sigma_{xx0}$. 
Consequently, the second term on the right-hand side $\left(\sigma_{xx0}^{-1} \sigma_{xx}-\sigma_{xx0}^{-2} \sigma_{xx}^2 \right)~\approx~0$~\cite{du_disorder-induced_2019_SI,huang_giant_2022_published_SI}.
This approximation allows us to simplify Eq.~\ref{ratio_3} scaling relation as,
\bea \label{ratio_4}
\Vxytwnormeq= \left({\mathcal C}_1 \sigma_{xx0}^{-1} +{\mathcal C}_2\sigma_{xx0}^{-2} - {\mathcal C}_4\sigma_{xx0}^{-2} \right)  \sigma_{xx}^2+   {\mathcal C}_4.
\eea
We use this simplified scaling relation for experimental fitting represented as
\be \label{eq:exptscaling}
\Vxytwnormeq = A \sigma_{xx}^2  +B~,
\ee
where $A=\left({\mathcal C}_1 \sigma_{xx0}^{-1} +{\mathcal C}_2\sigma_{xx0}^{-2} - {\mathcal C}_4\sigma_{xx0}^{-2}\right)={\mathcal C}^{sk2}_{0} \sigma_{xx0}^{-1}+\left({\mathcal C}^{sj}_0+ {\mathcal C}^{sk1}_{00}-{\mathcal C}^{sj}_1- {\mathcal C}^{sk1}_{11}\right)\sigma_{xx0}^{-2}$ represents the slope and $B={\mathcal C}_4$ is the intercept.
The intercept contains the BCD contribution, whereas the slope is solely determined by the skew-scattering and side-jump mechanism.

\subsection{$D/\epsilon_0$ as a parameter to probe the scaling relation}
One can perform the scaling analysis in Eq.~\ref{eq:exptscaling} by varying experimentally controllable parameters such as temperature, perpendicular electric field, etc.
In general, the scattering mechanisms, and hence, the scaling parameters $A$ and $B$ can be a function of temperature~\cite{xiao_scaling_2019_SI,du_disorder-induced_2019_SI}.
We used the perpendicular electric field \D{} as the parameter, at a fixed temperature, and performed the fitting within a suitable range of \D{} where the BCD and scattering contributions remain relatively constant. 
This scaling relation allowed us to capture the BCD dependence on the polarity of \D{} from the intercept $\eta$ of \Exytwnorm=($L^2$/$w$)\Vxytwnorm=$\zeta \sigma_{xx}^2  +\eta$.
Recent experiments report the observation of this scaling relation in different systems and use this analysis to estimate the order of magnitude of BCD, such as in few-layer WTe$_2$~\cite{kang_nonlinear_2019_SI}, where BCD$\sim \eta E_F/e$.
Here, $E_F$ is the Fermi energy and is fixed experimentally by the filling factor $\nu$ to probe the scaling relation.
To obtain the BCD dependence on the polarity of \D{}, the scaling is performed for a similar |\D{}| range for both $\pm$ \D{} polarities, and the intercept is compared.

In TDBG, when the perpendicular electric field is varied over a large \D{} range, the bands undergo valley Chern transitions (that is, a change in valley Chern numbers) that change the Berry curvature dipole.
Therefore, we further performed the local intercept analysis as discussed below,  and analyze the variation of the intercept over a large \D{} range.

\subsection{Local Intercept analysis} 
We considered a small interval ($D_1/\epsilon_0,\ D_2/\epsilon_0$) within which the BCD and scattering contributions can be considered unchanged.
A straight line was drawn connecting ($\frac{E_{xy}^{2\omega}}{(E_{xx}^\omega)^2}, \sigma_{xx}^2$)$|_{(D_1/\epsilon_0)}$ and ($\frac{E_{xy}^{2\omega}}{(E_{xx}^\omega)^2}, \sigma_{xx}^2$)$|_{(D_2/\epsilon_0)}$ to extract the local intercept $\eta(D_1/\epsilon_0)$. 
This approach can be visualized as finding the tangent at $D_1/\epsilon_0$. 
The local intercept of the tangent was then analyzed as a function of $D/\epsilon_0$, with an interval, $D_1/\epsilon_0-\ D_2/\epsilon_0 = 0.017$ \Vbynm{} (0.013 \Vbynm{}) as shown in Fig.~\ref{fig:SI_Lintercept_ABAB1} (Fig.~\ref{fig:SI_Lintercept_ABBA}) at additional filling factor $\nu$ aside the one presented in the main manuscript, near the CNP for 1.43$^{\circ}$ AB-AB (1.4$^{\circ}$ AB-BA) TDBG.
This method provides additional understanding of the nonlinear Hall voltage in TDBG by accounting for the dynamic changes in band structure when the \D{} is varied substantially over a large range.

\section{Additional scaling data from AB-AB TDBG devices}

\subsection{Scaling data from AB-AB TDBG device-1 with twist angle 1.43$\degree$}\label{sec:ABABdev1}
\begin{figure*}[h]
    \centering
    \includegraphics[width=16cm]{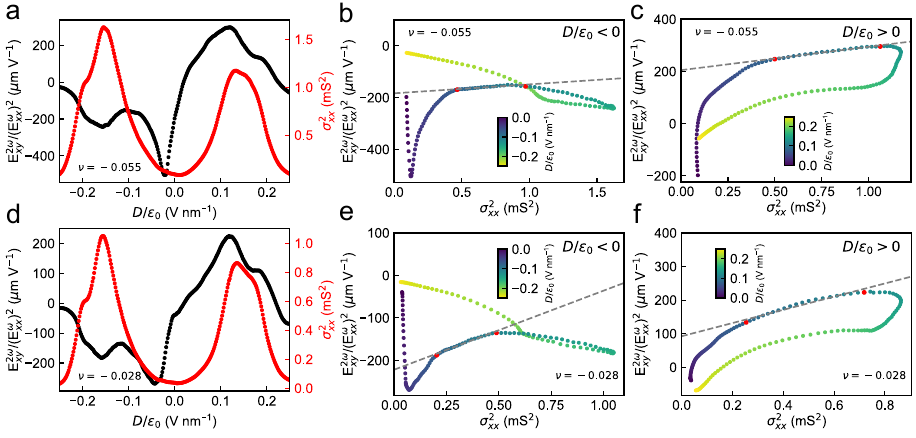}
    \caption{ \label{fig:SI_Gintercept_ABAB_additional_nu}
    \textbf{Scaling of normalized nonlinear Hall signal \Exytw/(\Exxw)$^2$ with the square of longitudinal conductivity ($\sigma_{xx}^2$) for different fillings $\nu$ in 1.43$^\circ$ AB-AB TDBG device.}
    \textbf{a-c,}~For filling $\nu=-0.055$ and \textbf{d-e,}~ for $\nu=-0.028$. \textbf{a,d,}~Variation of normalized nonlinear Hall signal \Exytw/(\Exxw)$^2$ (black colored data points corresponding to the left axis) and square of longitudinal conductivity $\sigma_{xx}^2$ (red-colored data points corresponding to the right axis) as a function of the displacement field $D/\epsilon_0$.
   \textbf{b,c,}~and \textbf{e,f,}~Parametric plots of the normalized nonlinear Hall signal \Exytw/(\Exxw)$^2$ against the square of the longitudinal conductivity  $\sigma_{xx}^2$ corresponding to data from \textbf{a} and \textbf{d} respectively, for \D{}<0 (\textbf{b,e}) and \D{}>0 (\textbf{c,f}).
   The color bar indicates the displacement field value of data points in \Vbynm{}.
   We note that the sign of the intercept changes for each of the fillings, as the polarity of \D{} is flipped.
   The measurements were performed using a current of 100~nA with a frequency of 177~Hz at a temperature of 1.2~K.
    }
\end{figure*}

\begin{figure*}[h]
    \centering
    \includegraphics[width=16cm]{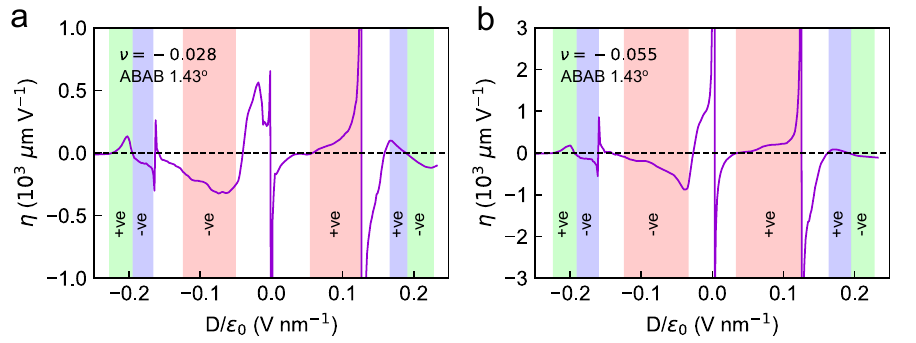}
    \caption{ \label{fig:SI_Lintercept_ABAB1}
    \textbf{Variation of local intercept over a large range of perpendicular electric field in AB-AB TDBG device-1 at different filling factor $\nu$.}
    \textbf{a, b,}~The local intercept $\eta$, defined for a small moving \D{} range of 0.017 \Vbynm{}, extracted as a function of \D{} from the data presented in figure~\ref{fig:SI_Gintercept_ABAB_additional_nu}d and ~\ref{fig:SI_Gintercept_ABAB_additional_nu}a for $\nu=-0.028$ (a) and $\nu=-0.055$ (b) respectively.
    The colors indicate the different \D{} ranges where $\eta$, and hence BCD, flip sign across $D=0$.
    The temperature was $T=1.2$~K.
    }
\end{figure*}

In this subsection, we probe the scaling relation in 1.43$\degree$ AB-AB TDBG for a few more filling factors, other than that shown in the main manuscript.
We show the \Exytwnorm{} and \sxxsq{} as a function of \D{} for $\nu=-0.055$ (Fig.~\ref{fig:SI_Gintercept_ABAB_additional_nu}a) and $\nu=-0.028$ (Fig.~\ref{fig:SI_Gintercept_ABAB_additional_nu}d).
The peak in \sxxsq{} indicates band touching as discussed in the main manuscript.
We probe the scaling of \Exytwnorm{} vs \sxxsq{} parametrically for both polarities of \D{} (Fig.~\ref{fig:SI_Gintercept_ABAB_additional_nu}b,c for $\nu=-0.055$ and Fig.~\ref{fig:SI_Gintercept_ABAB_additional_nu}e,f for $\nu=-0.028$), in the \D{} range just before the bands touch, similar to our analysis in Fig.~3d ($-$ve \D) and Fig.~3e (+ve \D) of the main manuscript.
We note that as the polarity of \D{} is flipped, the intercept reverses sign for both the fillings.
To understand the behavior of the parametric plots for the full \D{} range probed, we plot the local intercept over the extended \D{} range in Fig.~\ref{fig:SI_Lintercept_ABAB1}a ($\nu=-0.028$) and Fig.~\ref{fig:SI_Lintercept_ABAB1}b ($\nu=-0.055$).
We note the characteristic reversal in the intercept $\eta$, as the polarity of \D{} is flipped, similar to that observed in Fig.~3f of the main manuscript.
These observations further suggest that the BCD in 1.43$^\circ$ AB-AB TDBG flips sign with the polarity of \D{}.

\FloatBarrier
\newpage
\subsection{Scaling data from AB-AB TDBG device-2 with twist angle 1.1$\degree$}\label{sec:ABABdev2}
\begin{figure*}[htbp]
    \centering
    \includegraphics[width=15cm]{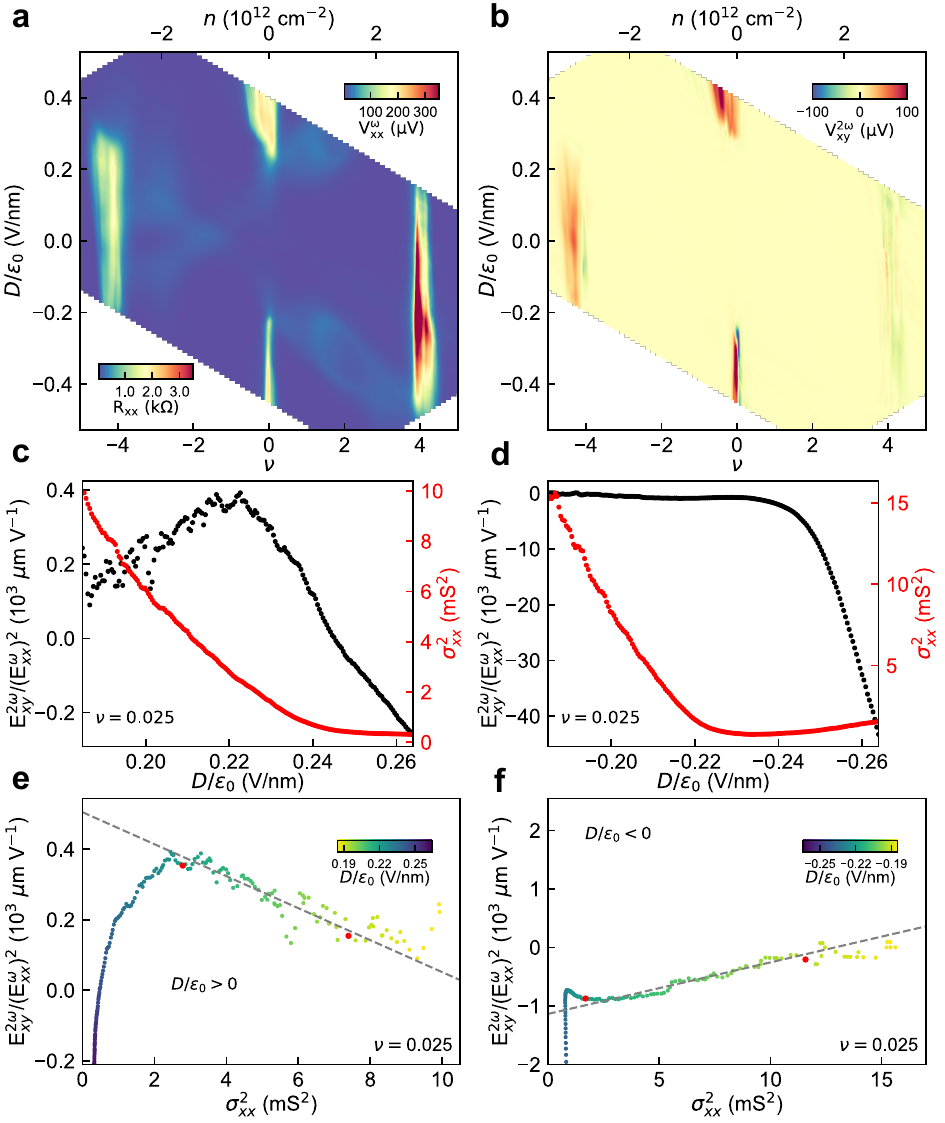}
    \caption{\footnotesize \label{fig:SI_intercept_ABAB2}
    \textbf{Scaling of normalized nonlinear Hall voltage \Vxytw/(\Vxxw)$^2$ with the square of longitudinal conductivity ($\sigma_{xx}^2$) in 1.1$^\circ$ AB-AB TDBG device-2.}
    \textbf{a,b,}~\Rxx{} (\textbf{a}) and NLH voltage \Vxytw{} (\textbf{b}) as a function of $\nu$ and \D{} of 1.1$^\circ$ AB-AB TDBG device. The top x-axis shows the corresponding $n$ values.
   \textbf{c,d,}~The variation of normalized nonlinear Hall signal \Exytw/(\Exxw)$^2$ (black colored data points corresponding to the left axis) and square of longitudinal conductivity $\sigma_{xx}^2$ (red-colored data points corresponding to the right axis) as a function of the displacement field $D/\epsilon_0$, for a fixed filling close to the charge neutrality gap.
   \textbf{e,f,}~The variation of normalized nonlinear Hall signal \Exytw/(\Exxw)$^2$ with square of longitudinal conductivity $\sigma_{xx}^2$ plotted parametrically as a function of the displacement field $D/\epsilon_0$, using \textbf{c}, and \textbf{d}, respectively, for \D{}>0 (\textbf{e}) and \D{}<0 (\textbf{f}).
   The color bar indicates the displacement field value of data points in \Vbynm{}.
   We note that the sign of the intercept flips, as the polarity of \D{} is flipped.
   The measurements were performed using a current of 100~nA with a frequency of 177~Hz at a temperature of 1.5~K.
    }
\end{figure*}

In Fig.~\ref{fig:SI_intercept_ABAB2}a and Fig.~\ref{fig:SI_intercept_ABAB2}b, we show the colorscale plot of \Rxx{} and the \Vxytw{}, respectively, as a function of $\nu$ and \D.
In Fig.~\ref{fig:SI_intercept_ABAB2}c and Fig.~\ref{fig:SI_intercept_ABAB2}d, we show the corresponding \Exytwnorm{}, \sxxsq{} vs. \D{} dependence with a fixed filling factor near the charge neutrality when the gap between the flat bands start opening~\cite{sinha_berry_2022_SI}, for positive and negative polarity of \D{} respectively.
Fig.~\ref{fig:SI_intercept_ABAB2}e and Fig.~\ref{fig:SI_intercept_ABAB2}f probe the corresponding scaling of \Exytwnorm{} with \sxxsq{} for positive and negative \D.
We note that the intercept flips sign as the polarity of \D{} is flipped.

\section{Additional scaling data from AB-BA TDBG device}
\begin{figure*}[h]
    \centering
    \includegraphics[width=16cm]{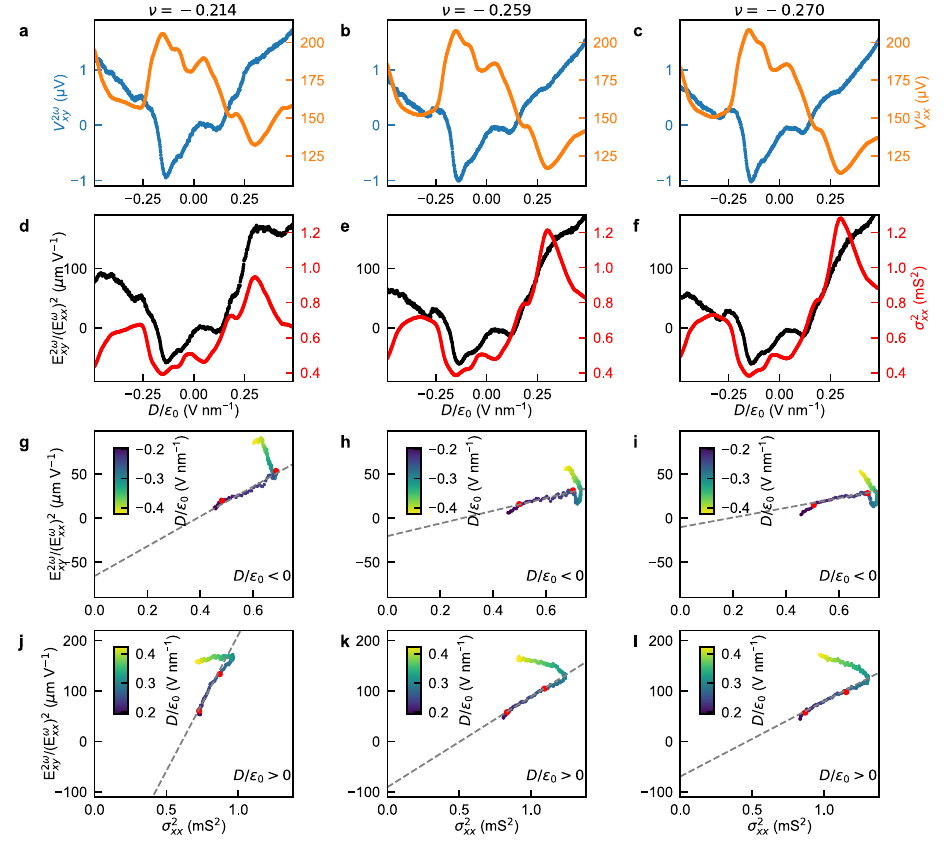}
    \caption{ \label{fig:SI_Gintercept_ABBA}
    \textbf{Scaling of normalized nonlinear Hall voltage \Vxytw/(\Vxxw)$^2$ with the square of longitudinal conductivity ($\sigma_{xx}^2$) for different fillings $\nu$ in 1.4$^\circ$ AB-BA TDBG device.}
    \textbf{a-c,}~The variation of nonlinear Hall voltage \Vxytw{} (blue-colored data points corresponding to the left axis) and longitudinal voltage \Vxxw{} (orange-colored data points corresponding to the right axis) as a function of the displacement field $D/\epsilon_0$ for three different fillings $\nu$ (different from those presented in Fig. 4 of the main manuscript).
   \textbf{d-f,}~The corresponding variation of normalized nonlinear Hall signal \Exytw/(\Exxw)$^2$ (black colored data points corresponding to the left axis) and square of longitudinal conductivity $\sigma_{xx}^2$ (red-colored data points corresponding to the right axis) as a function of the displacement field $D/\epsilon_0$, extracted for the same fillings used in \textbf{a}, \textbf{b}, and \textbf{c}, respectively.
   \textbf{g-l,}~The variation of normalized nonlinear Hall signal \Exytw/(\Exxw)$^2$ with square of longitudinal conductivity $\sigma_{xx}^2$ plotted parametrically as a function of the displacement field $D/\epsilon_0$, using \textbf{d}, \textbf{e}, and \textbf{f}, respectively, for \D{}<0 (\textbf{g-i}) and \D{}>0 (\textbf{j-l}).
   The color bar indicates the displacement field value of data points in \Vbynm{}.
   We note that the sign of the intercept remains the same for each of the fillings, as the polarity of \D{} is flipped.
   The measurements were performed using a current of 100~nA with a frequency of 177~Hz at a temperature of 1.5~K.
    }
\end{figure*}

\begin{figure*}[h]
    \centering
    \includegraphics[width=16cm]{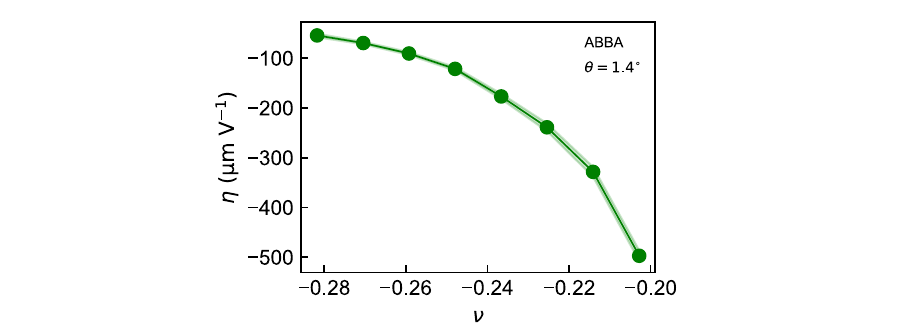}
    \caption{ \label{fig:intercept_nu_ABBA}
    Extracted intercept $\eta$ as a function of filling factor ($\nu$) for the valence band and positive perpendicular electric field ($D/\epsilon_0$) in AB-BA TDBG with a twist angle of 1.4$^\circ$. The BCD ($\Lambda$) $\sim \eta E_F/e$, decreases in magnitude as the filling factor moves away from the band edge. Individual fittings are shown in Fig.~\ref{fig:SI_Gintercept_ABBA} 
    }
\end{figure*}

\begin{figure*}[h]
    \centering
    \includegraphics[width=16cm]{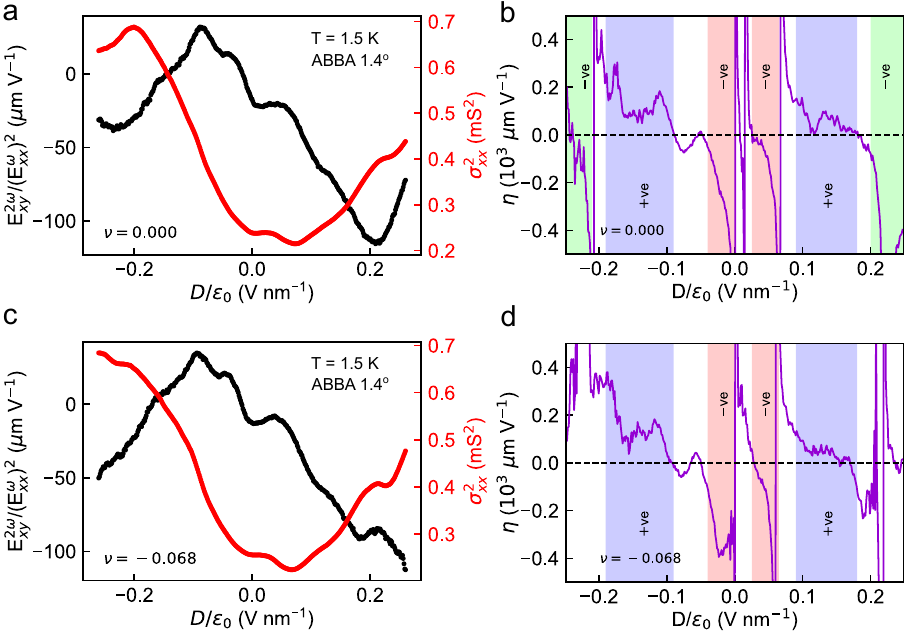}
    \caption{ \label{fig:SI_Lintercept_ABBA}
    \textbf{Variation of local intercept over a large range of perpendicular electric field in AB-BA TDBG device at different filling factor $\nu$.} \textbf{a,c,}~Variation of normalized nonlinear Hall signal \Exytw/(\Exxw)$^2$ (black colored data points corresponding to the left axis) and square of longitudinal conductivity $\sigma_{xx}^2$ (red-colored data points corresponding to the right axis) as a function of the displacement field $D/\epsilon_0$ for $\nu=0.0$ (a) and $\nu=-0.068$ (c).
    \textbf{b, d,}~The local intercept $\eta$, defined for a small moving \D{} range of 0.013 \Vbynm{}, extracted as a function of \D{} from the data in \textbf{a} and \textbf{c} respectively.
    The colors indicate the different \D{} ranges where $\eta$, and hence BCD, has same sign across $D=0$.
    The temperature was $T=1.5$~K.
    }
\end{figure*}

In this section, we probe the \D{} polarity dependence of the scaling relation for a few more filling factors other than that shown in the main manuscript, for the 1.4$^\circ$ AB-BA TDBG device.
Fig.~\ref{fig:SI_Gintercept_ABBA}a, b, and c shows the measured nonlinear Hall voltage \Vxytw{} and longitudinal voltage \Vxxw{} as a function of \D{} for three different $\nu=-0.214, -0.259, -0.270$.
Using the measured \Vxxw{}  and \Vxytw{}, we plot the corresponding normalized NLH electric field \Exytwnorm{} and $\sigma_{xx}^2=$~(I/\Vxxw)$^2$ as a function of \D{} for the three fillings in Fig.~\ref{fig:SI_Gintercept_ABBA}d, e, and f.
In Fig.~\ref{fig:SI_Gintercept_ABBA}g-l, we plot the \Exytwnorm{} vs. $\sigma_{xx}^2$ parametrically with \D{}, for \D{}<0 (Fig.~\ref{fig:SI_Gintercept_ABBA}g-i) and \D{}>0 (Fig.~\ref{fig:SI_Gintercept_ABBA}j-l).
We probe the scaling for the low \D{} range just before the conductivity maximizes, for both polarities of \D{}.
We observe a linear scaling in a similar range of |\D{}| range for both the polarities of \D.
We observe that the sign of the intercept ($\eta$) remains the same as the polarity of \D{} is flipped, and the extracted magnitude of $\eta$ decreases as one moves away from the band edge into the valence band (Fig.~\ref{fig:intercept_nu_ABBA}).
This BCD dependence is consistent with that extracted in Fig. 4d of Zhong, J. et al.~\cite{zhong_effective_2024_SI}.
In Fig.~\ref{fig:SI_Lintercept_ABBA}b ($\nu=0$) and Fig.~\ref{fig:SI_Lintercept_ABBA}d ($\nu=-0.068$), we further show the local intercept over the extended \D{} range for two fillings close to the charge neutrality gap.
We note that the intercept $\eta$ does not flip the sign in similar \D{} ranges indicated by the same color, as the polarity of \D{} is flipped, like that observed in Fig.~4f of the main manuscript.
These observations further suggest that the BCD in 1.4$^\circ$ AB-BA TDBG device does not flip sign with the polarity of \D{}.

\begin{figure*}[h]
    \centering
    \includegraphics[width=16cm]{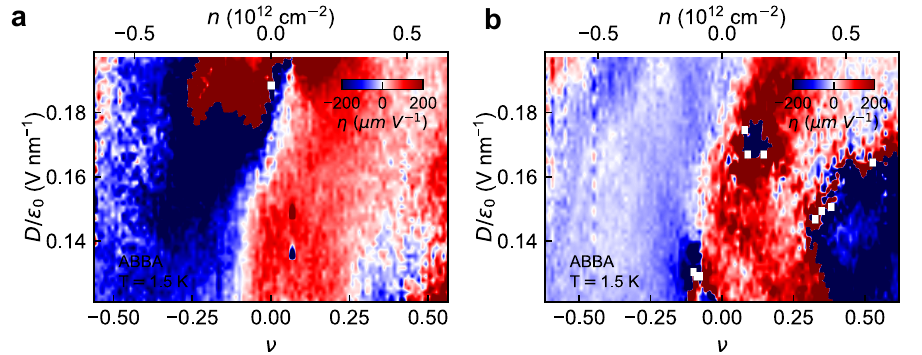}
    \caption{ \label{fig:local_intercept_nu_D_ABBA}
    Extracted intercept $\eta$ as a function of filling factor ($\nu$) and the perpendicular electric field ($D/\epsilon_0$)  for AB-BA TDBG with a twist angle of $\sim$1.4$^\circ$ for two different pairs of probe combinations. The BCD ($\Lambda$) $\sim \eta E_F/e$, flips sign across $\nu=0$, that is, as one moves from the valence band to the conduction band. This sign flip is consistent with the theoretically calculated BCD sign flip in Fig. 4a of the main manuscript across E=8 meV.
    }
\end{figure*}

\end{document}